\documentclass[a4paper,11pt]{article}
\pdfoutput=1  
\usepackage[utf8]{inputenc}
\usepackage{jheppub}

\usepackage{amssymb}
\usepackage{physics}
\usepackage{graphicx}
\usepackage{hyperref}
\usepackage{slashed}
\usepackage{orcidlink}
\usepackage{bbm}

\newcommand{\be}{\begin{equation}}
\newcommand{\ee}{\end{equation}}
\newcommand{\bea}{\begin{eqnarray}}
\newcommand{\eea}{\end{eqnarray}}

\newcommand{\z}[1]{\mathbb{Z}_{#1}}

\begin{document}
 \title{Constraints on discrete global symmetries in quantum gravity}
 \author[a]{Passant Ali\orcidlink{0000-0002-4468-9968},}
 \author[b]{Astrid Eichhorn\orcidlink{0000-0003-4458-1495},}
 \author[c]{Martin Pauly\orcidlink{0000-0002-7737-6204},}
 \author[a]{Michael M. Scherer\orcidlink{0000-0003-0766-9949}}

 \affiliation[a]{Institute for Theoretical Physics, University of Cologne, 50937 Cologne, Germany}
 \affiliation[b]{CP3-Origins,  University  of  Southern  Denmark,  Campusvej  55,  DK-5230  Odense  M,  Denmark}
 \affiliation[c]{Institut f\"ur Theoretische Physik, Ruprecht-Karls-Universit\"at Heidelberg, \\
Philosophenweg 16, 69120 Heidelberg, Germany}

\emailAdd{ali@thp.uni-koeln.de}
\emailAdd{eichhorn@cp3.sdu.dk}
\emailAdd{m.pauly@thphys.uni-heidelberg.de}
\emailAdd{scherer@thp.uni-koeln.de}

\abstract{
The question whether global symmetries can be realized in quantum-gravity-matter-systems has far-reaching phenomenological consequences. 
Here, we collect evidence that within an asymptotically safe context, discrete global symmetries of the form $\z{n}$, $n>4$, cannot be realized in a near-perturbative regime.
In contrast, an effective-field-theory approach to quantum gravity might feature such symmetries, providing a mechanism to generate mass hierarchies in the infrared without the need for  additional fine-tuning.
}

\maketitle

\section{Introduction}
Symmetries lie at the core of every quantum field theory (QFT). 
They restrict possible interactions and thereby shape the physical properties of QFTs. 
For example, in effective theories for solid state physics, a plethora of global symmetries is realized. 
Here, the interplay of continuous and discrete symmetries  contributes to the rich phenomenology on these scales. 
Zooming in towards more microscopic scales, this picture changes. 
At particle physics scales, continuous symmetries which can be gauged are predominant. 
In particular, the Standard Model (SM) of particle physics does not realize discrete internal symmetries beyond simple reflections.
This observation triggers the immediate question, whether there is a fundamental mechanism that prohibits discrete internal symmetries.

One possible  answer  could come from the interplay of the SM with quantum gravity. 
This interplay shapes the properties of nature at the most fundamental level:
the structures of the SM arise from --- or at least are constrained by --- the consistency of a microscopic matter-gravity model. 
Within an approach to quantum gravity, one could hence aim to identify which symmetries are permitted.

One example of this idea is given by recent progress on asymptotically safe gravity-matter models, where the constraining power of asymptotic safety might narrow down the allowed field content~\cite{Dona:2013qba,Meibohm:2015twa,Biemans:2017zca,Christiansen:2017cxa,Alkofer:2018fxj,Bonanno:2018gck,Wetterich:2019zdo,Daas:2020dyo}, values of couplings~\cite{Harst:2011zx,Eichhorn:2017eht,Eichhorn:2017lry,Eichhorn:2019dhg,Eichhorn:2020kca}, and masses~\cite{Shaposhnikov:2009pv,Eichhorn:2017ylw,Eichhorn:2018whv,Reichert:2019car,Kwapisz:2019wrl,Eichhorn:2020kca,Kowalska:2020gie} as well as dimensionality~\cite{Eichhorn:2019yzm}, see \cite{Eichhorn:2017egq,Eichhorn:2018yfc,Reuter:2019byg,Pawlowski:2020qer,Bonanno:2020bil} for reviews.
Another prominent example is the swampland program in string theory that aims at delineating the boundary between theories that could arise from a string theory and those that could not, see~\cite{Palti:2019pca} for a review. 
Within this context, arguments against the existence of global continuous symmetries exist~\cite{Banks:1988yz,Kamionkowski:1992mf,Kallosh:1995hi,Banks:2010zn} and have been substantiated to some degree~\cite{Harlow:2018jwu,Harlow:2018tng,Harlow:2020bee} in AdS/CFT. In their simplest form, these arguments rely on the incompatibility of the conservation of a global charge and a complete Hawking evaporation of black-hole solutions that would follow by an extension of semi-classical arguments into the quantum-gravity regime.
Along a similar line, arguments related to a finite lifetime of a de Sitter geometry have been invoked to argue against the realization of spontaneously broken discrete symmetries~\cite{Dvali:2018txx}.

In this paper, we explore whether discrete symmetries could be part of a quantum-field theoretic and ultraviolet (UV) complete description of nature within the asymptotic-safety paradigm.
Asymptotic safety is the realization of quantum scale symmetry in the UV, a.k.a.~an interacting fixed point of the Renormalization Group (RG) flow. 
In contrast to classical scale symmetry, quantum scale symmetry \cite{Wetterich:2019qzx} holds in the presence of quantum fluctuations, which naturally induce a scale dependence in the dynamics, expressed in scale-dependent couplings. 
Realizing quantum scale symmetry therefore requires a delicate balance between various interactions that constrains the  UV regime. 
Even once a non-vanishing RG flow sets in, the microscopically realized quantum scale symmetry leaves its imprints and leads to predictive power for the infrared (IR) regime.
For instance, there are several known examples where quantum scale symmetry at microscopic scales prohibits a non-vanishing  value for a specific coupling in the IR, e.g.,~\cite{Shaposhnikov:2009pv,Eichhorn:2017als,Pawlowski:2018ixd,Eichhorn:2020sbo}. 
These IR constraints on interaction structures are like the Cheshire cat's smile of asymptotic safety or quantum scale symmetry: 
In the IR, the symmetry is no longer realized, but it leaves its imprints through constraints on the values of couplings. 
This gives rise to the constraints on matter content, masses, couplings, and potentially even dimensionality, as mentioned above.
Here, we expand these results by exploring whether and why fundamental QFTs might prefer continuous  symmetries over discrete ones.

To do so, we focus on a particular subclass, namely  $\z{n}$ symmetric interactions and  investigate two questions:
\begin{enumerate}
  \item[1)] Can an interacting fixed point feature a global $\z{n}$ symmetry in the presence of quantum gravity?
  \item[2)] Can $\z{n}$ symmetric interactions appear along the RG flow as relevant perturbations of a gravity-matter fixed point with a larger (continuous) symmetry group?
\end{enumerate}
Answering  these allows to decide whether or not a discrete symmetry can be realized in the IR in a model that is asymptotically safe in the UV.

Within the scenario that we consider here, we work with a complex scalar field that naturally features a U(1) global symmetry if all $\z{n}$ symmetric interactions are set to zero.
We first consider a purely scalar system and then extend this system by a fermion charged under the global $U(1)$ to explore the impact of fermionic fluctuations on our results.
The spontaneous breaking of the U(1) symmetry is associated with an energy scale which sets the mass scale for the massive mode. 
In order to separate this mass scale from the cutoff scale of the theory, i.e., the scale of new physics, a fine-tuning of the initial conditions for the RG flow of the mass parameter is required.
The explicit breaking of the U(1) to a $\z{n}$ sets a second mass scale, associated with a mass for the pseudo-Goldstone mode. 
A large hierarchy between these two scales arises ``naturally" without any additional fine-tuning in this setting \cite{Leonard:2018sbi}.
We will discuss this mechanism in more detail in Sec.~\ref{sec:mass_hierarchy} and will explore its embedding into a UV completion to answer the question
\begin{enumerate}
  \item[3)] Can $\z{n}$ symmetric interactions lead to a large separation of energy scales in an asymptotically safe and approximately  U(1)-symmetric theory?
\end{enumerate}
To tackle these three questions, we apply the functional renormalization group (FRG) that we review in Sec.~\ref{sec:method}. Additionally, we discuss symmetries in this context. 
In Sec.~\ref{sec:qg_effects} we then introduce quantum gravitational effects into the RG flow. With these preconditions we first study discrete symmetries within asymptotic safety in Sec.~\ref{sec:single_field}. 
We then proceed to explore the generation of mass hierarchies in more general effective-field theory settings in~Sec.~\ref{sec:mass_hierarchy}.

\section{Setup and Method}
\label{sec:method}

\subsection{Functional Renormalization Group}

Using the FRG, we integrate out quantum fluctuations momentum shell by momentum shell. 
On the one hand, this allows to construct the effective dynamics from a given underlying microscopic dynamics. 
On the other hand, it allows to search for points in the space of couplings where the integration of momentum shells does not trigger a change in the couplings, i.e., RG fixed points.
This is implemented by considering the scale-dependent effective action $\Gamma_k$, a generalization of the effective action that takes into account all quantum fluctuations with momenta $q^2>k^2$.  
The scale-dependent effective action is obtained by introducing a mass-like regulator $R_k$ that suppresses fluctuations with $q^2<k^2$.
The scale dependence of $\Gamma_k$ is given by the flow equation~\cite{Wetterich:1992yh,Ellwanger:1993mw,Morris:1993qb}
\be
  \label{eq:flow_eq}
  \partial_t \Gamma_k = \frac{1}{2} \Tr \left( \partial_t R_k \left(\Gamma_k^{(2)} + R_k \right)^{-1}\right)\,,
\ee
with $t = \log(k/k_0)$ and $k_0$ a reference scale. 
Here, $\Gamma_k^{(2)}$ is the second variation of $\Gamma_k$ with respect to the fluctuating fields. 
The trace symbolizes a trace in field space, as well as in momentum space and over spacetime and any internal indices. 
For Grassmann-valued fields it acquires an additional minus sign. 
For $k\to 0 $, all fluctuations are integrated out and one obtains the effective action, $\Gamma = \Gamma_{k\to 0}$. 

For gravity, the use of this framework has been pioneered in the seminal work of M.~Reuter \cite{Reuter:1996cp}, leading to substantial evidence for asymptotic safety in gravity, see, e.g., \cite{Reuter:2001ag,Litim:2003vp,Codello:2008vh,Benedetti:2009rx,Manrique:2010am,Manrique:2011jc,Falls:2013bv,Christiansen:2014raa,Becker:2014qya,Christiansen:2015rva,Gies:2016con,Denz:2016qks,Bosma:2019aiu,Knorr:2019atm,Falls:2020qhj}, see \cite{Percacci:2017fkn,Eichhorn:2018yfc,Reuter:2019byg,Pereira:2019dbn,Eichhorn:2020mte,Reichert:2020mja,Bonanno:2020bil,Pawlowski:2020qer} for recent reviews and lecture notes.

While \eqref{eq:flow_eq} is an exact equation, in practice one needs to truncate the effective action $\Gamma_k$ to a subset of all operators compatible with the symmetries. 
By projecting onto these operators, one can extract the scale dependence of the couplings, i.e., their beta functions, from \eqref{eq:flow_eq}. 
For  reviews of the method in the context of various fields of physics, see~\cite{Pawlowski:2005xe,Gies:2006wv,Delamotte:2007pf,Dupuis:2020fhh}.

Consider the beta function for a dimensionless coupling $g = k^{-d_{\bar{g}}} \bar{g}$ with $\bar{g}$ the dimensionful counterpart and $d_{\bar{g}}$ its canonical scaling dimension. 
It can be written as
\be
  \label{eq:beta_1}
  \beta_{g} = - d_{\bar{g}}\, g + b(g),
\ee
with $b(g)$ a term induced by interactions that is typically of higher order in the couplings. 
Neglecting this term, the scale dependence of dimensionless couplings is hence
\be
  \label{eq:canonical_scaling}
  g = g(k_0)\left( \frac{k}{k_0}\right)^{-d_{\bar{g}}}
.\ee
Accordingly, canonically relevant (irrelevant) couplings increase (decrease) towards the IR. 
This is still true in a perturbative regime, where the scaling dimensions receive corrections from quantum fluctuations, but the signs of quantum scaling dimensions typically agree with the canonical dimension. 
This distinction will be key for the phenomenological implications in Sec.~\ref{sec:mass_hierarchy}.
Finally, for canonically marginal couplings with $d_{\bar{g}}=0$, quantum fluctuations can render them marginally relevant or marginally irrelevant.

\subsection{Implementation of Asymptotic Safety}

Asymptotic safety provides well-defined initial conditions in the microscopic limit, i.e., as $k\to \infty$. 
These are provided by a fixed point $g_\ast$, $\eval{\beta_{g}}_{g=g_\ast} = 0$.
Towards the IR, deviations from the fixed point can either grow or shrink. They grow/shrink if the critical exponent
\be
  \theta_i = -\text{Eig}_i\left(\pdv{\beta_{g_j}}{g_l} \right)
\ee
is positive/negative. 
The corresponding eigendirection is called a relevant/irrelevant direction. 
The value of every irrelevant superposition of couplings\footnote{At an interacting fixed point, the eigenvectors of the stability matrix $\pdv{\beta_{g_j}}{g_l}$ are typically not aligned with the original couplings $g_i$, instead it is typically superpositions of couplings which are relevant/irrelevant. As long as the fixed point is near-perturbative in nature, this mixing is typically negligible.} 
is fixed when flowing towards the IR. 
Along such directions, quantum fluctuations act to restore quantum scale symmetry.
Conversely, quantum fluctuations act to increase the breaking of quantum scale symmetry along the relevant directions, for which the deviation from the fixed-point value can in principle grow large.
Thus, the values of relevant couplings need to be fixed by measuring them at a reference scale $k_0$.

If an eigendirection is approximately aligned with a coupling $g$ and interactions are negligible, then the corresponding critical exponent is determined by the  canonical scaling dimension of $g$, as apparent from~\eqref{eq:beta_1}. 
Accordingly, as long as the system is in a near-perturbative regime, the canonical scaling dimension of operators provides guidance on how to set up robust truncations.
Within the context of asymptotically safe gravity-matter systems, the assumption of near-perturbativity is supported by numerous results in the purely gravitational sector \cite{Falls:2013bv,Falls:2014tra,Denz:2016qks,Falls:2017lst,Falls:2018ylp,Kluth:2020bdv}, as well as in the interplay of gravity and matter \cite{Eichhorn:2017eht,Eichhorn:2017sok,Eichhorn:2018akn,Eichhorn:2018nda,Eichhorn:2018ydy,Eichhorn:2020sbo}. 
In particular, in gravity-matter systems, the coupling of sufficiently many fermion and vector fields could drive the system into a more perturbative regime \cite{Dona:2013qba, Eichhorn:2018nda}.

\subsection{RG Flow and Symmetries}
\label{sec:flow_symm}

The space of all couplings corresponding to all possible interactions, unrestricted by any global symmetries, constitutes the most general theory space. 
Each global symmetry defines a hypersurface in theory space.
Unless an explicit breaking of the symmetry is introduced by the regulator, the flow preserves the global symmetry and therefore remains in the symmetric hypersurface, once it starts within that surface.\footnote{Local symmetries differ in that they are typically violated by the introduction of the regulator, thereby the flow does not remain in the symmetric hypersurface and symmetry identities have to be imposed, see, e.g., \cite{Gies:2006wv,Pawlowski:2020qer}.}
This follows, as the most symmetric form of the propagator $(\Gamma_k^{(2)}+R_k)^{-1}$ which drives the flow, see Eq.~\eqref{eq:flow_eq}, exhibits all symmetries of the kinetic term.

One may break the symmetries of the kinetic term by means of additional interactions, i.e., by allowing initial conditions for the RG flow outside the symmetry-enhanced hypersurface. 
Consider the additional couplings $g_i', g_j', \dots$ corresponding to a group $G'$ which is a subgroup of the symmetry group $G$ of the kinetic term.\footnote{Here, we assume that no additional kinetic terms are compatible with $G'$. 
This is the case for the $\z{n}, n \geq 3$ symmetries we consider below.} In other words, the corresponding interactions break the original symmetry group $G$ down to one of its subgroups.
For instance, the propagator for a complex scalar field respects a global U(1) symmetry, which is not respected by the coupling of a $\z{3}$ symmetric interaction $\phi^3+(\phi^{\ast})^3$.
The corresponding beta function for $g_i'$ necessarily vanishes if all $g_i', g_j', \dots$ are set to zero, i.e., the flow does not break the global symmetry $G$. Thus the flow of symmetry-breaking couplings can only be nonzero if at least one symmetry-breaking coupling is nonzero.

A nontrivial zero of the flow could arise at finite values of the couplings $g_i', g_j', \dots$ .
Such an additional fixed point would feature the reduced symmetry $G'$.
A simple example of such a case for a scalar field would be a fixed point at finite $\phi^4$ coupling, breaking the shift symmetry of the kinetic term.
If such an interacting fixed point does not exist, this prevents $G'$ from being realized (in a non-trivial way) in the UV. 
The only remaining fixed-point will be the Gau\ss{}ian fixed point of $G'$, $g_i'= g_j'=\dots=0$.

For the RG flow towards the IR, there are two possibilities: 
i) If there is a relevant coupling associated to deviations from that fixed point, the theory could still exhibit the smaller symmetry $G'$ in the IR instead of the full symmetry $G$. 
ii) If the Gaussian fixed point only features irrelevant directions  outside the hypersurface defined by $G$, even the IR theory is prevented from exhibiting the symmetry $G'$ within an asymptotically safe theory.
This would put the symmetry into the ``asymptotically-safe swampland", as it cannot emerge from an asymptotically safe UV fixed point. 

\section{Quantum Gravitational Effects}
\label{sec:qg_effects}

We will now focus on the case of a complex scalar field $\phi$, which features a global U(1) symmetry. 
All U(1) symmetric interactions are denoted by $\bar{\lambda}_i$, and their dimensionless counterparts by $\lambda_i$. 
Additionally, we consider interactions which explicitly break the global U(1) symmetry to a global $\z{n}$ symmetry. 
These interactions are parameterized by the couplings $\bar{z}_i$ and their dimensionless counterparts $z_i$. To explore the impact of fermionic fluctuations on the scalar potential, we also consider a Dirac fermion coupled to the complex scalar via a Yukawa coupling $y$.

Within asymptotic safety, the breaking of a global U(1) symmetry to a global $\z{n}$ symmetry in the UV would require an interacting fixed point for the couplings in the $\z{n}$ sector. 
A breaking of the global U(1) to a global $\z{n}$ symmetry in the IR is compatible with a non-interacting fixed point, but only if a $\z{n}$ symmetric coupling is associated to a relevant direction.
To study whether any of these two scenarios can arise at trans-Planckian scales, one needs to take into account gravitational fluctuations. 
In the following, we discuss the contributions of these fluctuations to the beta functions of scalar couplings.

\subsection{Structure of Gravitational Contributions}

\begin{figure}
\centering
 \includegraphics{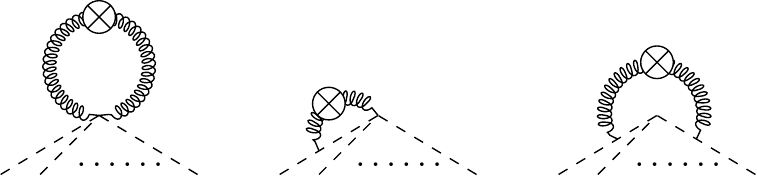}
 \caption{
 \label{fig:feynman_grav}
 We show the three diagrams involving gravitational fluctuations (curly lines) that form the dominant gravitational contributions to the beta function of an $n$-point scalar coupling. The second and third diagram also come in the form where the regulator insertion sits on the internal scalar lines.
 }
\end{figure}

The three generic gravitational contributions to the beta function of a momentum-independent scalar $n$-point coupling $z_n$ are depicted in Fig.~\ref{fig:feynman_grav}. 
All of them are proportional
to $z_n$. This is a direct consequence of the one-loop structure of the flow equation, together with the fact that the flow equation preserves global symmetries.
They can hence be accounted for by introducing a term\footnote{Note that different conventions for the sign of this gravitational contribution are also chosen in the literature.}
\be
\Delta \beta_{z_n} = - f_s(\Lambda, g) z_n,\label{eq:gravcontr}
\ee
into the beta function for $z_n$. Thus, 
\be
\beta_{z_n}^\text{grav} = \beta_{z_n}^\text{w/o\, grav} + \Delta\beta_{z_n},
\ee
is the full beta function including gravitational fluctuations. 
The gravitational contribution effectively acts as a gravity-induced anomalous dimension $f_s(\Lambda,g)$ which depends on the fixed-point values of the dimensionless version of the cosmological constant $\Lambda$ and the Newton coupling $g$ and further gravitational couplings. 

In a heuristic picture, one can think of quantum fluctuations of spacetime as changing the ``effective" spacetime dimensionality that matter fields ``experience".
In the simplest approximation, $f_s$ is independent of additional couplings in the matter sector.
The term $f_s$ does not depend on $n$ nor the symmetry structure of the coupling, i.e., it is the same for a $U(1)$-breaking coupling $z_n$ and a $U(1)$ symmetric coupling $\lambda_n$. 
This is a consequence of the fact that the gravitational field is ``blind" to the internal symmetries that matter fields exhibit.
In fact, the same gravitational contribution is at the heart of a proposed mechanism that makes the ratio of the Higgs mass to the electroweak scale calculable from first principles in asymptotic safety \cite{Shaposhnikov:2009pv}.

In each sector defined by its own symmetry $G'$, Eq.~\eqref{eq:gravcontr} holds for the coupling of the lowest-order interaction.
For higher-order couplings, additional gravitational contributions arise which are proportional to the lower-order couplings in the $G'$-sector. 
For our argument, it is sufficient to focus on the lowest-order couplings.

We neglect non-minimal couplings to gravity, such as $\xi \phi\phi^\ast R$, with $R$ the Ricci scalar. 
This does not affect our main argument regarding $\z{n}$ symmetric couplings. More specifically, in the absence of fermionic fluctuations, such non-minimal couplings feature a vanishing fixed-point value, see, e.g.,  \cite{Narain:2009fy,Eichhorn:2020sbo}. In the presence of fermionic fluctuations and finite Yukawa couplings, such a non-minimal coupling can be present, but does not affect the sign of $f_s$ \cite{Eichhorn:2020sbo}. Additionally, non-minimal couplings that respect the $\mathbb{Z}_n$ symmetry vanish, as long as we focus on fixed points at which the U(1) symmetry is unbroken at the fixed point, as we do for the main body of the paper. For the studies in the appendix, it is important to note that such non-minimal couplings are of canonically higher order (even compared to the $z_n$-coupling) and could therefore be expected to be subleading, at least at near-perturbative fixed points.

Similarly, we parameterize the contribution of gravitational fluctuations to the beta function of Yukawa couplings by a factor $f_y$. 
The beta function for the Yukawa coupling then reads  \cite{Zerf:2017zqi,Oda:2015sma,Eichhorn:2016esv, Eichhorn:2017eht,Hamada:2017rvn,Eichhorn:2017ylw,deBrito:2019epw}
\be
  \label{eq:beta_yukawa}
  \beta_y = \frac{1}{4 \pi ^2} y^3 - f_y(\lambda, g) y 
.\ee
The part cubic in the Yukawa is the universal one-loop result.

The contributions $f_y$ and $f_s$ are typically obtained in Euclidean signature. 
While one can relate these results to the Lorentzian ones via an analytical continuation on a fixed background with an appropriate Killing vector, in the context of gravity this procedure is more subtle, see, e.g., \cite{Baldazzi:2018mtl}. 
We assume that the Euclidean computations can be applied to obtain a qualitative understanding of the Lorentzian theory. 

\subsection{Effect of Gravitational Contributions}
\label{sec:grav_contribs}
The two quantities $f_s$ and $f_y$ parameterize the influence of gravitational fluctuations on matter couplings. 
Within asymptotic safety both depend on the fixed-point values for the gravitational couplings. 
In turn, the fixed-point values for $(\Lambda, g)$ and higher-order gravitational couplings depend on the matter content of the theory \cite{Dona:2013qba, Meibohm:2015twa, Eichhorn:2016vvy, Biemans:2017zca,Christiansen:2017cxa,Eichhorn:2018nda,Alkofer:2018fxj,Wetterich:2019zdo,Burger:2019upn}. 
Additionally, in current state-of-the-art computations, they depend on a variety of technical choices, see, e.g., \cite{Gies:2015tca,Ohta:2016npm,deBrito:2018jxt} that leave a relatively large theoretical uncertainty on $f_s$ and $f_y$, see, e.g., \cite{Eichhorn:2016esv}.
We do not assume specific values here. 
Instead we vary $f_s$ and $f_y$ to parameterize the impact of gravitational fluctuations and explore which dynamics is accessible in the quantum-gravity regime.

While the magnitude of $f_s$ depends on the gravitational fixed-point values \cite{Eichhorn:2017als,Pawlowski:2018ixd,deBrito:2019umw,Wetterich:2019rsn, Eichhorn:2020sbo}, see also \cite{Narain:2009fy,Percacci:2015wwa,Labus:2015ska,Wetterich:2016uxm,Hamada:2017rvn} for works which evaluate it at a given gravitational fixed point, its sign is generally found to be  negative. 
Evaluated at the Gau\ss{}ian fixed point for matter couplings, the resulting positive linear contribution \eqref{eq:gravcontr} drives scalar couplings towards irrelevance.
Beyond the Gau\ss{}ian matter fixed point at finite gravitational but vanishing matter couplings, no other fixed points have been found in pure scalar-gravity systems \cite{Narain:2009fy,Narain:2009gb,Percacci:2015wwa,Labus:2015ska,Wetterich:2016uxm, Eichhorn:2017als,Pawlowski:2018ixd, deBrito:2019umw}.

In an appropriate regime of the quantum gravitational parameter space, $f_y > 0$ might be realized \cite{Eichhorn:2016esv, Eichhorn:2017eht, deBrito:2019umw,Eichhorn:2020sbo}, triggering the emergence of an interacting fixed point in the Yukawa coupling, cf.~Eq.~\eqref{eq:beta_yukawa}. 
At this interacting fixed point, $y$ is irrelevant. 
This structure might even allow for a computation of the top mass within the asymptotic-safety framework \cite{Eichhorn:2017eht,Eichhorn:2017ylw,Eichhorn:2018whv,Alkofer:2020vtb}. 
In turn, the non-vanishing value of the Yukawa coupling breaks the shift symmetry associated with the Gau\ss{}ian fixed point and induces a nontrivial scalar potential \cite{Eichhorn:2019dhg,Eichhorn:2020kca,Eichhorn:2020sbo}.

Consequently, there are two scenarios:
(i)~if  all $\z{n}$-symmetric matter-couplings vanish in the UV, only relevant couplings can trigger deviations from the $U(1)$ symmetry in the IR.
(ii)~Alternatively the $\z{n}$-symmetric sector itself might feature an interacting fixed point that breaks shift symmetry.
We will discuss both scenarios in the following.

\section{Discrete Symmetries in Asymptotic Safety}
\label{sec:single_field}

We first focus on models that contain a single complex scalar field $\phi$ and then explore how our conclusions are affected in models with more than one scalar field.
We consider a complex scalar field $\phi$ with the (scale-dependent) effective action
\be
  \label{eq:action_u1}
  \Gamma^{\mathrm{U(1)}}_k = \int \dd[4]{x} \sqrt{g} \left( Z_\phi g^{\mu\nu} \partial_\mu \phi^\ast \partial_\nu \phi + V(\phi,\phi^\ast) \right)
.\ee
The potential 
\be
V(\phi, \phi^\ast) = V_{\mathrm{U(1)}}(\phi\phi^\ast) + V_{\z{n}}(\phi, \phi^\ast),
\ee
is split into a U(1) invariant part $V_{\mathrm{U(1})}$ and a part $V_{\z{n}}$ that implements the discrete symmetry~\cite{Leonard:2015wyg,Classen:2017hwp,Zinati:2017hdy,Leonard:2018sbi,Torres:2018jij,Jian:2018aqk,Codello:2020mnt}. 
We additionally introduce a Dirac fermion $\psi$ with the effective action
\be
  \Gamma_k^\text{ferm} =\!\int\!\dd[4]{x}\!\sqrt{g} \left( i \bar{\psi} \slashed{\nabla} \psi + y \left( \phi^\ast \bar{\psi}_R \psi_L\!-\! \phi \bar{\psi}_L \psi_R \right) \right)
,\ee
where the subscripts $R$ and $L$ mark the right- and left-handed part $\psi_{L/R} = P_{L/R} \psi$ and $P_{L/R} = \frac{1}{2}\left(1 \pm \gamma_5 \right)$. 
For more details on our conventions in the fermionic sector, see, e.g.,~\cite{Gies:2001nw,Gies:2009sv}. While fermionic fluctuations are not key for the arguments presented in this section, due to the negative sign of their contribution in the beta function for the scalar mass, they drive the scalar potential towards a symmetry-broken regime under the RG flow to the IR, as is well known also from the Standard Model \cite{Eichhorn:2015kea}.
This will become important in the analysis of phenomenological consequences of discrete symmetries in Sec.~\ref{sec:mass_hierarchy}. For this reason we choose to include the Yukawa coupling whenever we present explicit results in this section and the appendix. 

We expand the U(1) invariant part of $V(\phi\phi^\ast)$ in terms of the U(1) invariant $\phi\phi^\ast$ as
\be
  V_{U(1)}\left(\phi \phi^\ast\right) = \sum_{i=1}^{i_\text{max}} \frac{\lambda_{2i}}{i!} (\phi\phi^\ast)^i.
\ee 
The additional interaction that we introduce breaks the global U(1) symmetry to a $\z{n}$ symmetry
\be
\label{eq:potential_zn}
V_{\z{n}}(\phi, \phi^\ast) = \bar{z}_n \left( \phi^n + (\phi^\ast)^n \right)
,\ee
with $\bar{z}_n$ a dimensionful coupling and canonical dimension
\begin{align}
    \left[\bar{z}_n\right] = d - n\,.
\end{align}
In addition, we consider a wave-function renormalization $Z_{\phi}$, leading to contributions from the anomalous dimension $\eta_\phi = - \partial_t \ln Z_\phi$ in the beta functions. 
The corresponding contributions to the beta function $\beta_{z_n}$ have the form $\eta_\phi z_n$ and  are hence important to decide whether or not a coupling is relevant at the free fixed point. 
Let us make a technical remark regarding $\eta_{\phi}$: 
We do not take into account the contributions from regulator derivatives in the numerator of the flow equation. 
These correspond to higher-order corrections and are expected to be negligible in the near-perturbative setting we explore.

As indicated before, the canonical relevance or irrelevance of the leading $\z{n}$-symmetric coupling $z_n$ is of central importance. 
We hence consider the canonically relevant case $n=3$, the marginal case $n=4$, and the irrelevant case $n>4$ separately. Before we go into more depth regarding the argument in each case, we briefly summarize our results.

For each case we first study the UV structure, i.e., whether it features an  interacting fixed point and whether this fixed point is near-perturbative. 
The latter requirement means that canonical scaling dimensions and quantum scaling dimensions are in approximate agreement. 
This is key for our truncation of the effective action to reliably capture the dynamics.
We then determine the (ir-)relevance of the corresponding couplings, and whether a generation of the $\z{n}$ symmetric interaction is possible along the RG flow.
 
We find that for $\z{3}$, an interacting fixed-point might exist  beyond the near-perturbative regime. In addition, the corresponding coupling is relevant at the Gaussian fixed point in a significant part of the gravitational parameter space.
The case $n=4$ is more constrained; whether it features a near-perturbative interacting fixed point is discussed in App.~\ref{app:nonpert}. It generically does not generate $\z{4}$ symmetric interactions towards the IR, starting from the free fixed point. 
The case $n>4$ neither features an interacting near-perturbative fixed point, nor relevant directions to depart from the Gau\ss{}ian fixed point. 
While this result holds in truncations, we will argue that it persists under an extension of the truncation.
 
For the fixed-point search, we will exclusively focus on the symmetric regime and work under the assumption that a fixed point in the symmetry-broken regime should show up in our truncations as a fixed point with negative mass parameter.

\subsection{General Case: \texorpdfstring{$\z{n}$}{Z\_n} with \texorpdfstring{$n>4$}{n>4}}

To explore whether there is an interacting fixed point, we study the beta function for a coupling $z_n$ in the $\z{n}$ symmetric sector. 
We focus on the near-perturbative case and comment on the general case in App.~\ref{app:nonpert}.
Our argument  relies on the following three observations:
\begin{enumerate}
 \item[i)] Already the lowest-order coupling $z_n$ in a given $\z{n}$ symmetric sector corresponds to an interaction that has $n$ external legs; therefore there cannot be a contribution $z_n^2$ in the beta function $\beta_{z_n}$ for $n>4$ due to the one-loop structure of the flow equation. In fact, all couplings in a $\z{n}$ symmetric sector with $n>4$ only occur linearly in their beta functions.\footnote{A diagram contributing to $\beta_{z_n}$ needs to have $n$ external legs. The identity $L=I-V+1$ with $L$, $I$ and $V$ the number of loops, internal legs and vertices, respectively,  reduces to $I = V$ for $L=1$, which is the flow-equation case. Accordingly, by adding one vertex, one also adds an internal line. Thus, an $n$-point vertex contributes $n-2$ external legs (notice that each internal line is connected to two vertices). For a diagram containing $i$ vertices proportional to $z_n$ we hence end up with (a minimum of) $(n-2)\cdot i$ external legs. But we require $(n-2) \cdot i \leq n$. This can be rewritten as $n \leq \frac{2i}{i-1}$. For $i \geq 2$ this requires $n \leq 4$. In particular $\beta_{z_{n}}$ does not contain any term quadratic or higher in $z_{n}$ for $n>4$. A similar argument applies for the anomalous dimension.}
\item[ii)] The lowest-order coupling $z_n$ in a given $\z{n}$ symmetric sector has negative canonical dimensionality for $n>4$, i.e., it is canonically irrelevant.
\item[iii)] Quantum gravitational contributions act like an anomalous scaling dimension for a given lowest-order $\z{n}$-symmetric coupling, i.e., they contribute linearly to the beta function of that coupling.\footnote{For higher-order couplings $z_i$ in a given $\z{n}$-symmetric sector, gravitational contributions proportional to (powers of) lower dimensional couplings contribute, if the corresponding couplings exist.} That scaling dimension is \emph{negative}, i.e., shifting couplings towards irrelevance, c.f. also Sec.~\ref{sec:grav_contribs}.
\end{enumerate}
Let us first provide a simple argument that focuses only on $z_n$ and neglects higher order couplings and then in a second step discuss how it generalizes beyond that simple approximation.

Due to points i) and iii), the beta function for $z_n$ is  linear in $z_n$, i.e.,
\be
\beta_{z_n}= -d_{\bar{z}_n}\, z_n - f_s\, z_n,
\ee
with the canonical dimension $d_{\bar{z}_n}<0$, since $z_n$ is canonically irrelevant. The only fixed-point solution is
\be
z_{n\, \ast}=0.
\ee
Canonically, $z_n$ is irrelevant at that fixed point, expressed in the negative critical exponent
\be
\theta_{\rm w/o\, grav} = d_{\bar{z}_n}<0.
\ee
This cannot be changed under the impact of quantum gravity, since the full critical exponent is
\be
\theta = d_{\bar{z}_n} + f_s <0,
\ee
since $f_s<0$ and $d_{\bar{z}_n}<0$.
Therefore, $z_n$ is not just zero at the UV fixed point, but has to remain zero at all scales.

Going beyond the simplest approximation for the $\z{n}$ symmetric sector, higher-order couplings exist which can feed back into $\beta_{z_n}$ (see, e.g., the coupling $z_{4,2}$ in Eqn.~\eqref{eq:beta_z4} below). 
As $z_n$ itself can appear in those beta functions, we can mimick their effect by higher-order $z_n^\alpha$, $\alpha\geq 2$ terms in $\beta_{z_n}$. 
Once such terms exist, a nontrivial zero of $\beta_{z_n}$ can arise as a consequence, depending on the sign of the higher-order term. 
Yet, such a zero can only exist if it was already present without the impact of quantum gravity, i.e., if four-dimensional scalar field theory on its own was asymptotically safe with an appropriate $\z{n}$ symmetric term. 
As we discuss in App.~\ref{app:nonpert}, we find no indications for the existence of such a fixed-point.

The argument presented above holds in $d\geq 4$ spacetime dimensions, as long as $f_s<0$ holds in $d\geq 4$ (see, e.g., App.~C of~\cite{Percacci:2015wwa}), since $d_{\bar{z}_n} = d+n-d \cdot n/2$.
Accordingly, for $d=4$, the case of $\z{4}$ is the case with a canonically marginal coupling; for all $n>4$, the above argument holds.
For $d=6$ and $f_s<0$, not only the leading $\z{4}$-symmetric interaction, but even the leading $\z{3}$ symmetric interaction are already canonically irrelevant.

\subsection{The \texorpdfstring{$\z{3}$}{Z\_3}-Symmetric Case}
\subsubsection{UV Fixed Point}
\label{sec:no_cubic_term}

In the $\z{3}$-symmetric case, we first focus on the fixed point at which $z_{3\, \ast}=0$, which is guaranteed to exist. 
As the gravitational sector is interacting, the critical exponents associated to the matter sector are not the canonical scaling dimensions, but acquire additional contributions from the interacting gravity sector. 
Specifically, the critical exponent associated to $z_3$ at $z_{3\, \ast}=0$ is 
\be
\theta_{z_3}= 1+f_s,
\ee
since $\eta_{\phi}=0$ at that fixed point within our truncation.
For $f_s>-1$, the coupling remains relevant. 
Accordingly, nonzero values of this coupling can be reached in the IR by the RG flow starting from a joint gravity-matter fixed point. 
In the region of gravitational parameter space in which $f_s>-1$, the existence of $\z{3}$ symmetric interactions is compatible with asymptotically safe gravity within our truncation. 

Additionally, a fixed point at $z_{3\, \ast}\neq 0$ is not a priori excluded. 
However, if it exists, it cannot be near-perturbative.
We do not consider such a fixed point in the main text and refer to App.~\ref{app:z3} for a brief discussion of this scenario where a fixed-point candidate is identified and the deviations of the scaling spectrum from perturbative scaling are calculated.

\subsubsection{Flow Towards the IR}

The presence of a $\z{3}$ symmetric coupling could lead to an intriguing IR phenomenology, which we briefly highlight here. 
If the U(1) symmetry is broken spontaneously in the absence of a $\z{3}$ coupling, a massless Goldstone boson is present. 
In the presence of a $\z{3}$ coupling, it acquires a mass proportional to the $z_3$ coupling.
In the spirit of the mechanism presented in detail in Sec.~\ref{sec:mass_hierarchy}, one could expect that in this case an inverted mass hierarchy might be realized, with the pseudo-Goldstone boson becoming more massive than the other massive mode. 
The underlying mechanism relies on canonical power counting: 
The coupling $z_3$ is canonically relevant. 
Thus, it grows towards the IR. 
If the vacuum expectation value is tuned towards criticality, such that the ratio between the Planck scale and the vev is large,
then naively one would expect the mass ratio between pseudo-Goldstone mode and massive mode to become large. 
However, this scenario is not realized in our truncation: 
As long as one only accounts for the coupling $z_3$, the mass ratio in the corresponding potential is bounded $M_\text{long}/M_\text{trans}>1/3$, with $M_{\rm long/trans}$ the two mass eigenvalues. 

In this respect, the scenario we consider here differs from the one considered in~\cite{Torres:2018jij}. In our case, the bounded mass ratio arises because $z_3$ contributes to both, the transversal as well as the longitudinal mass. In~\cite{Torres:2018jij} the effective potential is constructed in field-space coordinates adapted to keep the minimum fixed in radial direction. This requires the inclusion of a non-analytic term $(\phi \phi^\ast)^{3/2}$. Such terms then might allow to generate an inverted mass hierarchy.  

Finally, we point out that  if the addition of a $\z{3}$-symmetric coupling $z_3$ circumvented the triviality problem of $\phi^4$ theory, $z_3$ would be irrelevant, see App.~\ref{app:z3} and the mass ratio between longitudinal and transverse modes would become calculable.

\subsection{The Case \texorpdfstring{$\z{4}$}{Z\_4}}
In the case $n=4$, we focus on the beta function for the coupling $z_4$ in the near-perturbative regime; for a discussion of the non-perturbative regime see App.~\ref{app:nonpert}.
The beta function for $z_4$ reads
\be
\label{eq:beta_z4}
\beta_{z_4} = (2\eta_\phi - f_s) z_4 + \frac{(6-\eta_\phi) \lambda_4 z_4}{8\pi^2 (1+\lambda_2)^3} - \frac{5 (6-\eta_\phi) z_{4,2}}{96\pi^2 (1+\lambda_2)^2}.
\ee
Here, $z_{4,2}$ is the coupling belonging to an interaction $(\phi \phi^\ast) (\phi^4+(\phi^\ast)^4)$.
Neglecting gravitational corrections, $f_s = 0$, the coefficient of the term linear in $z_4$ remains positive as long as $\lambda_4$ remains positive. 
A near-perturbative fixed point could arise if $z_4$ would become relevant at the (potentially shifted) Gau\ss{}ian fixed point, or correspondingly the linear coefficient would change sign.
We find that this is not the case:
\begin{enumerate}
  \item[(i)] Without gravitational corrections, we do not discover a near-perturbative interacting fixed point, see App.~\ref{app:nonpert}.
  \item[(ii)] Gravitational corrections are known to shift scalar couplings towards irrelevance, $f_s < 0$. They hence strengthen the irrelevance of the coupling $z_4$ at the free fixed point.
\end{enumerate}
Hence, it is impossible to generate a finite $z_4$ by flowing towards the IR. 
Therefore, our results indicate that in the IR a $\z{4}$ symmetry is not realized in an asymptotically safe theory in the near-perturbative regime and instead, the larger U(1) symmetry remains unbroken.

\subsection{Extension to Two-Field Models}
\label{sec:multi_field}

One can extend the results from the previous subsections to the case with more than one field. If each of the fields is charged with a charge of one under the $\z{n}$ group, the discussion of the previous subsections apply.

If various fields are charged differently, more elaborate constructions are possible. 
In App.~\ref{app:two_fields} we discuss an example, that implements the following simple idea: 
Consider a $\z{6}$ symmetry and a scalar $\phi$ with charge $1$ under that symmetry. 
If one adds a second scalar $\chi$, charged with charge $-2$ under this symmetry, the second scalar is also charged under the $\z{3}$ subgroup of the $\z{6}$. 
By the arguments of the last section, all couplings in the $\phi$ sector vanish at a fixed point, as long as the two sectors are not coupled.

The situation changes if one introduces a portal from the $\phi$ to the $\chi$ sector. 
Non-vanishing coupling values in the $\chi$ sector can generate interactions in the $\phi$ sector. 
In particular, this induces the $\z{6}$ symmetry in the $\phi$ sector.
More generally, this mechanism bypasses some of the results of the last section by transferring the  explicit $U(1)$ violation in terms of an effective $\z{3}$ symmetry from the $\chi$ sector to a $\z{3q}$ sector by means of a portal, where $q$ is the charge of the scalar $\chi$.
We highlight that the corresponding interacting fixed point is likely not of near-perturbative nature as it features deviations in the critical exponents from the canonical scaling dimension that are $\order{1}$ (cf.~Eqn.~\eqref{eqn:app_two_fields_crit_exps}).

Due to the significant technical complexity, we do not compute the beta functions in the symmetry broken regime and do not explore the flow towards the IR, but merely point out that multi-field scenarios could provide a possible mechanism to circumvent the results in the single-field case we have presented in the last subsections.

\section{Discrete Symmetries in the Effective-Field-Theory Approach to Quantum Gravity}
\label{sec:mass_hierarchy}

Let us now broaden our point of view beyond asymptotic safety. 
In fact, any predictive UV completion could leave a testable imprint in the IR by fixing the initial conditions for the RG flow in the UV, and thereby the value of IR parameters.
Generically, this happens in two steps: 
Beyond a  momentum scale $k_\text{UV}$, physics is governed by the UV completion, and may not necessarily be formulated as a local quantum field theory. 
Below $k_\text{UV}$, a quantum field theoretic regime must set in. 
This is necessary in order to connect the UV completion to known physics in the IR, given that all experimental data on particle physics are well-describable within a quantum field theory.
The initial conditions for the RG flow of the couplings in the QFT at the scale $k_\text{UV}$ are determined by the UV completion. 
Depending on how predictive the UV completion is, there might be a range of initial conditions or just a single value that is compatible with the UV completion.
The mapping from $k_\text{UV}$ into the IR then proceeds via the standard RG flow.

In this section we remain agnostic about the particular UV completion. 
Instead, we take an effective field theory point of view and explore which phenomenology can arise from given initial conditions at $k_\text{UV}$. We choose $k_{\rm UV} \geq M_{\rm Planck}$. Motivated by the conjecture that quantum gravitational effects could break global symmetries, at least in some selected quantum-gravity approaches, we assume that the U(1) symmetry is not protected in the quantum gravitational regime. Accordingly, we assume finite values of the $\z{n}$ symmetric interactions at $k_{\rm UV}$ and explore the resulting phenomenology.
In short, the scaling \eqref{eq:canonical_scaling} can induce large mass hierarchies in a natural way.
In this section, we discuss this mechanism and 
(i)~study the structure of the relevant effective potential, (ii)~explore how the RG flow generates a mass hierarchy in a natural way and (iii)~investigate how quantum gravitational fluctuations impact the $\z{n}$ symmetric couplings which lie at the heart of the hierarchy-generating mechanism. Lastly, we provide a specific example to illustrate this mechanism.

%------------------------------
\subsection{Classical Analysis}
%------------------------------

\begin{figure}
    \centering
    \includegraphics[width=0.6\textwidth]{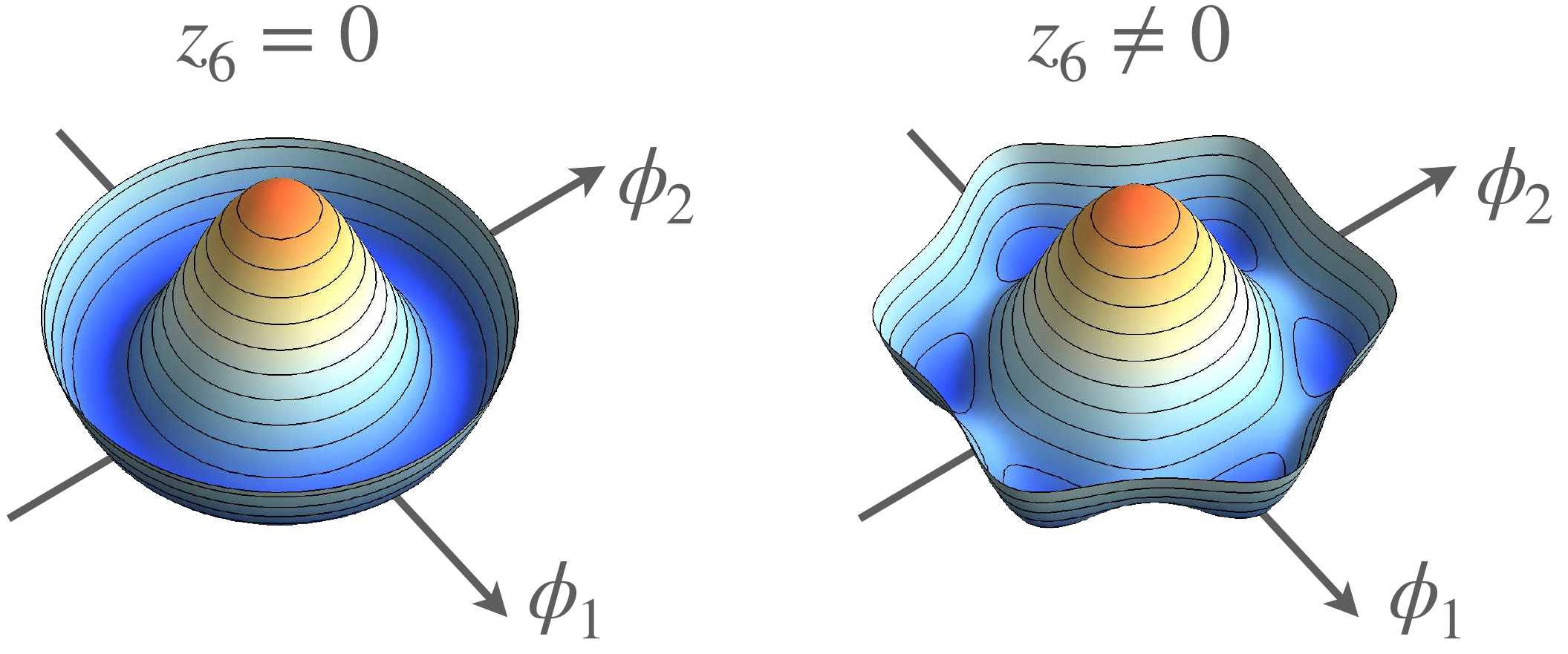}
    \caption{Potential with vanishing/non-vanishing $\z{6}$ coupling $z_6$ on the left/right. The massless $U(1)$ Goldstone mode acquires a mass, once the $U(1)$ symmetry is broken to a $\z{6}$ symmetry.
    We plot the potential for $\phi_1, \phi_2$ with $\phi = \frac{1}{\sqrt{2}} (\phi_1 + i \phi_2)$.
    }
    \label{fig:potential_z6}
\end{figure}

In order to understand the vacuum structure of the  effective scalar potential, we first neglect the $\z{n}$ interaction and consider a purely U(1) invariant potential
%. 
%
in the spontaneously symmetry-broken regime, 
where we employ an expansion of the effective potential around a non-vanishing minimum $\bar \kappa$, i.e.,
\begin{align}\label{eq:u1pot}
  V_{U(1)}(\phi \phi^\ast) = \frac{\lambda_{4}}{2} (\phi\phi^\ast- \bar{\kappa})^2\,.
\end{align}
 In general, quantum fluctuations also generate higher-order terms in the effective potential.
For simplicity, we only consider the lowest-order term here  which suffices for our analysis. 
This potential is sketched in the left panel of Fig.~\ref{fig:potential_z6}. 
Excitations around the minimum in the radial direction acquire a finite mass
\be\label{eq:u1mass}
M_\text{long}=\sqrt{2\bar{\kappa} \lambda_4}\,.
\ee
In contrast, excitations in the transversal (angular) direction remain massless, $M_\text{trans}=0$, corresponding to the Goldstone mode.

U(1)-breaking $\z{n}$-symmetric interactions, cf.~Eq.~\eqref{eq:potential_zn}, destroy the flatness of the potential along the angular direction.
In consequence, the explicit U(1) symmetry breaking by the $\z{n}$ symmetric interaction turns the Goldstone boson into a pseudo-Goldstone boson. 
Its mass is set by the symmetry-breaking coupling. To derive the mass, we redefine the $\z{n}$ symmetric interaction introduced in  Eq.~\eqref{eq:potential_zn} as
 \be
 \label{eq:rep_Vn}
    \hspace{-.17em} V_{\z{n}}(\phi, \phi^\ast) = \bar{z}_n
     \left(-\left[\phi^n + (\phi^\ast)^n\right] + 2(-1)^n {(\phi \phi^\ast)}^{\frac{n}{2}} \right)
 .\ee
This redefinition introduces non-analyticities for odd $n$.
In the following, we exclusively consider the case of even $n$, with $n>4$.
The full effective potential with the U(1) breaking term from Eq.~\eqref{eq:rep_Vn} acquires $n$ degenerate minima, cf.~right panel in Fig.~\ref{fig:potential_z6}. In the limit $z_n \rightarrow 0$, or also $n\rightarrow \infty$, 
these merge to form the familiar U(1)-manifold of degenerate minima.

%-------------
%-------------

 The redefinition of the potential according to Eq.~\eqref{eq:rep_Vn} by adding the term $\propto (\phi \phi^\ast)^{\frac{n}{2}}$ 
 %ensures that for any choice of $n$, at least one minimum is aligned with the $\phi_1$-axis, such that $\phi_2\eval_{\text{min}}=0$. Further, for even $n$, it 
 implies that the position of the minimum along the $\phi_1$ direction is located at $\phi_1\eval_{\text{min}}=\pm\sqrt{2\bar{\kappa}}$ as in the U(1) case, cf.~Eq.~\eqref{eq:u1pot}.
Without loss of generality we focus on the minimum at $(\phi_1,\phi_2)=(-\sqrt{2\bar{\kappa}}\ ,0)$ in the following. This redefinition 
is convenient as it allows to disentangle the effect of the $z_n$ coupling onto to the two massive modes, as also done in \cite{Torres:2018jij}.

To obtain the masses, we diagonalize the mass matrix, i.e., the matrix of second derivatives of the action with respect to the two fields, evaluated at constant field values, around that minimum. The masses are given by
\bea
  \label{eq:masses_zn}
  M_{\text{long}}^2 &=&  2\bar{\kappa} \lambda_4\,,\\
  M_{\text{trans}}^2 &=&  n^2 \bar{z}_n \bar{\kappa}^{\frac{n}{2}-1}\,,\\
  M_{\text{ferm}}^2 &=& y^2 \bar{\kappa}
  \label{eq:Mtrans}
.\eea
The transversal mass is proportional to $\bar{z}_n$ and, hence, it does not vanish anymore, once $\bar{z}_n$ is finite. 
 Note that due to the parametrization chosen in Eq.~\eqref{eq:rep_Vn}, the form of the longitudinal mass does not change as compared to the U(1) case\footnote{This is easiest to see by rewriting $\phi=\phi_1 + i \phi_2$, with $\phi_{1,2}$ real fields. For example, for the case of $n=6$, the redefined potential in Eq.~\eqref{eq:rep_Vn} is then simply a rotation by $\pi/12$ in $
\phi_{1,2}$ of $V_{\z{6}}(\phi_1, \phi_2)= \bar{z}_6
{\left(\phi_1-\phi_2\right)^2}\left(\phi_1^2+4\phi_1 \phi_2 +\phi_2^2 \right)^2$ used in \cite{Leonard:2018sbi}.}, cf.~Eq.~\eqref{eq:u1mass}.
The (dimensionless) ratio of the scalar masses is
\be
  \label{eq:mass_ratio}
  \gamma \equiv \frac{M_{\rm trans}^2}{M_{\rm long}^2}
  = \frac{n^2 \bar{z}_n \bar{\kappa}^{\frac{n}{2}-2}}{2\lambda_4}
  =\frac{n^2 z_n \kappa^{\frac{n}{2}-2}}{2\lambda_4},
\ee
where we converted to dimensionless quantities in the last step.
In the next subsection we will illustrate how the RG flow can render this quantity small. 

%-------------------------------------------
\subsection{Mass Hierarchy Due to RG Running}
%-------------------------------------------

To exemplify the generation of a mass hierarchy, we use the numerical values describing the Standard Model Higgs field. We note, however, that the mechanism we explore here cannot straightforwardly be extended to a Higgs field that carries charge under a gauge group as opposed to a global symmetry. We comment on potential phenomenological applications of our study in the conclusions.

As our model contains both bosonic as well as fermionic quantum fluctuations, initial conditions for the RG flow can be chosen such that the scalar potential is driven into a symmetry-broken regime. 
In particular, the mass parameter for the Higgs field which sets the electroweak scale is a relevant parameter.\footnote{This holds below the Planck scale. 
In the quantum gravitational regime the situation can change, if quantum gravitational fluctuations are strong enough, see, e.g., \cite{Wetterich:2016uxm,Eichhorn:2017als,Pawlowski:2018ixd,Eichhorn:2020sbo}.}
Its smallness with respect to $M_\text{Pl}$ signals the near-criticality of the Standard Model; one needs to fine-tune the UV value of the Higgs mass parameter at $k_{\rm UV}$ such that over 16 orders of magnitude one stays sufficiently close to the critical value $\kappa_{\rm crit}$.\footnote{We argue in terms of $\kappa$ here, which is of course negative in the symmetric regime, where the RG flow is typically expressed in terms of a coupling $\lambda_2$ instead. For the sake of our argument, this technical difference does not play a role.} 
While this does not constitute a problem of consistency, discontent with this enormous tuning is commonly expressed as the ``hierarchy problem'' of the Standard Model. 

\begin{figure}
    \centering
    \includegraphics{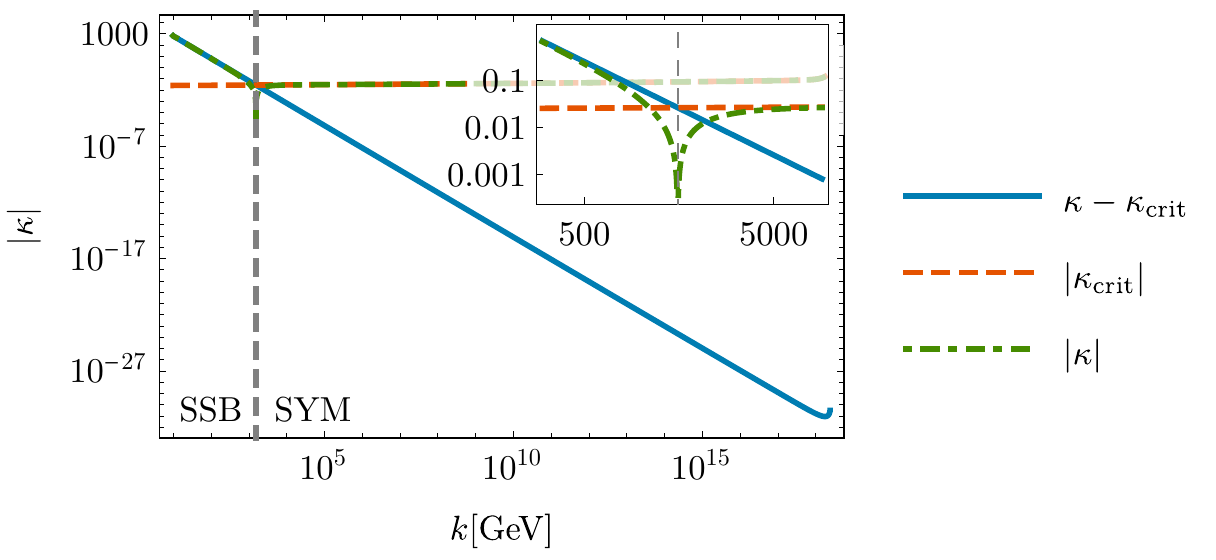}
    \caption{Running of the mass parameter $\kappa$ with RG scale $k$ for $n=6$. For large energies $\kappa$ is dominated by $\kappa_\text{crit}$, for small energies the deviation from the critical hyper-surface dominates.  The dashed vertical line marks the onset of the symmetry-broken regime. The initial conditions chosen here are those described in the caption of Fig.~\ref{fig:masses_no_grav}.}
    \label{fig:deviation_from_crit}
\end{figure}

Within our model we do not tackle this issue, but take such a tuning as given and instead focus on the natural emergence of a second hierarchy:
Once $\kappa$ is tuned close to criticality, an additional hierarchy between $M_{\rm long}$ and $M_{\rm trans}$ follows automatically, as explained in the following.
We comment on the case $k_{\rm UV}> M_{\rm Planck}$ below and focus on $k_{\rm UV} = M_{\rm Planck}$ first.
Based on its canonical dimensionality, the Higgs mass parameter, or equivalently, $\kappa$, generically increases with $k^2$ under the RG flow to the IR. 
This results in a Higgs mass close to the Planck scale, if the dimensionless mass is chosen of order 1 at the cutoff scale, in contradiction with observations. To accommodate the observed Higgs mass, $M_{\rm long}^2/M_{\rm Planck}^2 \sim 10^{-34}$, the initial condition for the RG flow of $\kappa$ must be tuned very closely to the critical point at which its flow vanishes.
Thus, the dominant scaling over a large range of scales is
\be
\kappa - \kappa_{\rm crit} = \delta \kappa \sim \left(\frac{k_{\rm UV}}{k}\right)^2,
\ee
where $\kappa_{\rm crit}$ is the critical value. The critical value itself is comparably large, cf.~Fig.~\ref{fig:deviation_from_crit}. 

\begin{figure}
    \centering
    \includegraphics{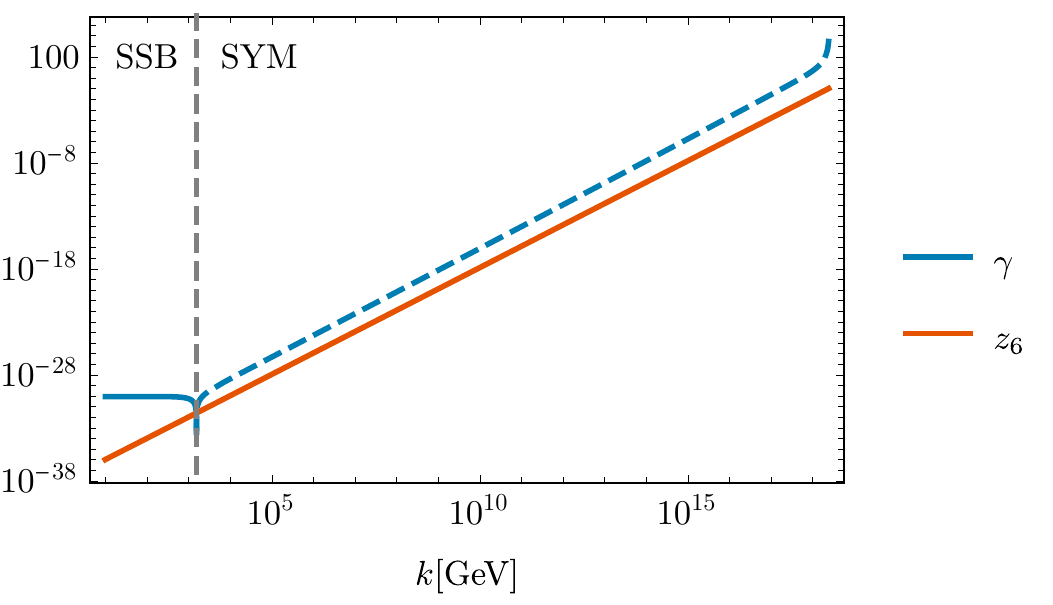}
    \caption{Running of the mass ratio $\gamma$  (cf.~\eqref{eq:mass_ratio}) as a function of the RG scale $k$ for $n=6$. In the symmetry-broken regime $\gamma$ corresponds to the ratio of the transversal and longitudinal masses. Towards the IR a  tiny mass ratio emerges. In the symmetric regime $\gamma$ is negative (dashed). The initial conditions chosen here are those described in the caption of Fig.~\ref{fig:masses_no_grav}.}
    \label{fig:gamma_running}
\end{figure}

Therefore, it is a good approximation to set $\kappa \sim \mathrm{const.}$ over a large range of scales, until the deviation from criticality has become appreciable.
We obtain Fig.~\ref{fig:deviation_from_crit} and the following Figures by fixing the initial conditions $y(M_{\rm Planck})=0.4, \lambda_4(M_{\rm Planck})=0.001, z_6(M_{\rm Planck})=0.1$ as exemplary values. %Working with high numerical precision we then fine-tuning 
$\kappa(M_{\rm Planck})$ is fine-tuned such that $\bar{\kappa} = \left(246 \text{GeV}\right)^2$ in the IR, we do not provide all digits of the expression here.

Quantum corrections to the scaling of $z_n$
can largely be neglected, and it scales according to
\bea
z_n &\sim& \left(\frac{k}{k_{\rm UV}} \right)^{n-4}.\label{eq:scalings}
\eea
Due to their vanishing canonical dimensionality, $\lambda_4$ and $y$ only depend logarithmically on $k$, i.e., compared to 
$z_n$, they are essentially constant.

Accordingly, all quantities except $z_n$
are approximately constant in the mass ratio $\gamma$, cf.~Eq.~\eqref{eq:mass_ratio}. The $k^{n-4}$ scaling of $z_n$ ensures that $\gamma$ is driven to tiny values with a power law, i.e.,
\be
\gamma \sim z_n \left(\kappa_{\rm crit}+\delta\kappa \right)^{\frac{n}{2}-2}.
\ee
We thus encounter two regimes: i) as long as $\kappa_{\rm crit}\gg \delta \kappa$, $\gamma \sim k^{n-4}$ due to the scaling of $z_n$; ii) once $\delta \kappa \geq \kappa_{\rm crit}$, $\gamma \sim \rm const$. Comparing Fig.~\ref{fig:deviation_from_crit} to Fig.~\ref{fig:gamma_running}, the transition between both regimes\footnote{Within a perturbative scheme in which $\kappa_{\rm crit}=0$, the first regime is absent. Instead, $\gamma$ is roughly constant and assumes its tiny value $\gamma \ll 1$ already at the Planck scale.} is clearly visible in $\kappa$ and consequently $\gamma$.
 
The IR value of $\gamma$ is smaller, the longer the above scaling holds, i.e., the longer $\kappa$ does not transition to its ``natural" scaling with $k^{-2}$, where $\gamma$ becomes constant. 
Thus, a tiny value of $\gamma$  emerges naturally without any further tuning, once $\kappa$ is tuned to ensure a small ratio $M_{\rm long}^2/M_{\rm Planck}^2 \sim 10^{-34}$. The ratio $M_{\rm trans}^2/M_{\rm long}^2 \sim \left(M_{\rm long}/M_{\rm Planck}\right)^{n-4}$ follows \emph{automatically}.

In summary, the presence of the discrete $\z{n}$ symmetry gives rise to an \emph{additional} mass hierarchy: If one fine-tunes the UV value of $\kappa$ such that $M_\text{long} \ll k_\text{UV}$, no further fine-tunging of $z_n$
is required to realize $M_\text{trans} \ll M_\text{long}$ in the IR. Instead the second mass hierarchy follows as a direct consequence of the RG running of $z_n$.
%-------------------------------------------
\subsubsection{Gravitational Enhancement of the Mass Hierarchy}
%-------------------------------------------

Let us investigate the situation in which $k_{\rm UV} > M_{\rm Planck}$. In that case, the quantum field theoretic, effective description contains quantum gravitational degrees of freedom. They should be understood as effective degrees of freedom that are applicable between $k_{\rm UV}$ and $M_{\rm Planck}$ and encode the gravitational microphysics that holds beyond $k_{\rm UV}$.

\begin{figure}
    \centering
    \includegraphics{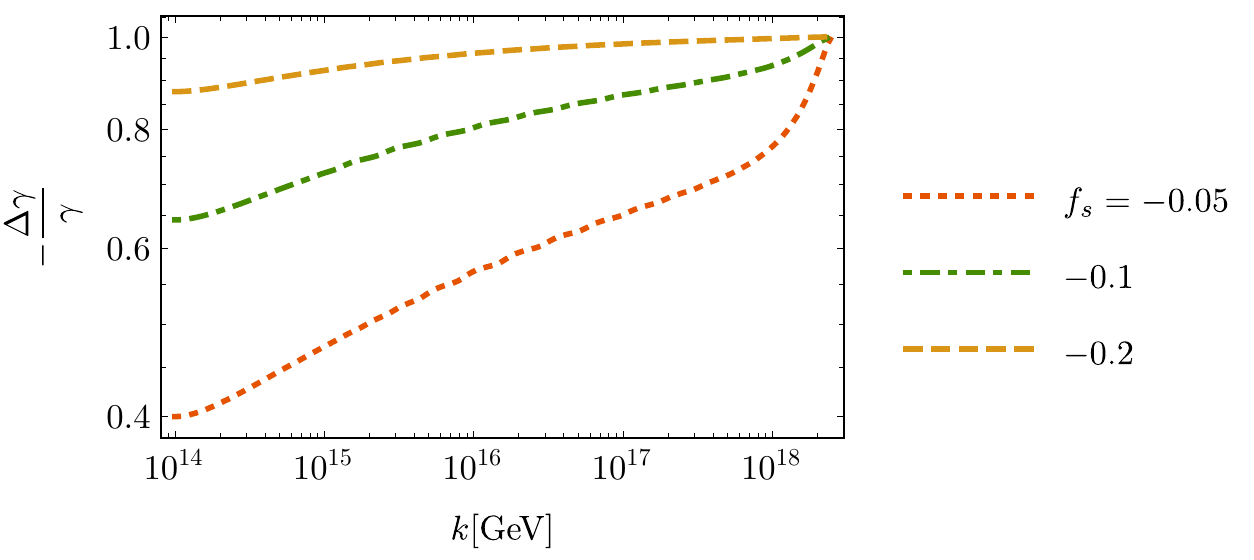}
    \caption{Difference $\frac{\Delta \gamma}{\gamma} = \frac{\gamma(f_s) - \gamma(f_s=0)}{\gamma(f_s=0)}$ in the mass ratio that arises due to the effect of gravitational quantum fluctuations, parameterized by $f_s$ for the case $n=6$. Motivated by Ref.~\cite{Eichhorn:2017ylw}, this plot uses a fiducial value of $f_y = 0.004$.}
    \label{fig:delta_gamma}
\end{figure}

In that situation, an anomalous scaling exponent determines the RG flow of the coupling $z_n$. Instead of Eq.~\eqref{eq:scalings}, in this range of scales, the following scaling holds:
\bea
z_n \sim \left(\frac{k}{k_{\rm UV}} \right)^{n-4 + f_s}.\label{eq:scalingQG}
\eea
In order to obtain the separation between $M_{\rm long}$ and $M_{\rm Planck}$, $\kappa$ now has to be fine-tuned at $k_{\rm UV}$ and a similar behavior to that discussed in the previous subsection holds. Without gravitational corrections, the scaling $\gamma \sim k^{n-4}$ holds. With gravitational corrections, the enhanced scaling $\gamma \sim k^{n-4+f_s}$ holds. Accordingly, quantum gravity enhances the resulting mass hierarchy even more, as can be seen in Fig.~\ref{fig:delta_gamma}. 
This can also be understood within the framework of effective asymptotic safety, where a quantitative measure of predictivity has been defined in \cite{Held:2020kze}.

In Fig.~\ref{fig:delta_gamma}, we set $n=6$ and artificially lower the Planck scale to $\tilde{M}_\text{Pl} = 10^{14}\, \text{GeV}$ simply for purposes of illustration, such that gravitational fluctuations to the flow contribute for $k > \tilde{M}_\text{Pl}$. 
We then parameterize the gravitational contributions by varying $f_s <0$. 
We keep $z_6\left(k = M_\text{Planck} \sim 10^{18} \text{GeV}\right)$ constant and additionally keep all other couplings at $\tilde{M}_\text{Pl} = 10^{14}\, \text{GeV}$ constant. The resulting mass ratio below $\tilde{M}_\text{Pl}$ becomes smaller due to the impact of gravitational fluctuations, when $|f_s|$ is increased. Gravitational fluctuations qualitatively act to enhance the mass hierarchy. As long as $|f_s| \ll n$, their quantitative influence is minor.

Fig.~\ref{fig:delta_gamma} also highlights that quantum gravity fluctuations drive the $\z{n}$ symmetric couplings towards zero. As $k_{\rm UV} \rightarrow \infty$, $ z_n(M_{\rm Planck}) \rightarrow 0$, resulting in the exclusion of $\z{n}$ $(n>4)$ symmetric interactions in asymptotically safe gravity.

%-------------------------------------------
\subsubsection{Concrete Example: \texorpdfstring{$\z{6}$}{Z\_6} Symmetry and Hierarchy-Generation}
%-------------------------------------------

\begin{figure}
    \centering
    \includegraphics{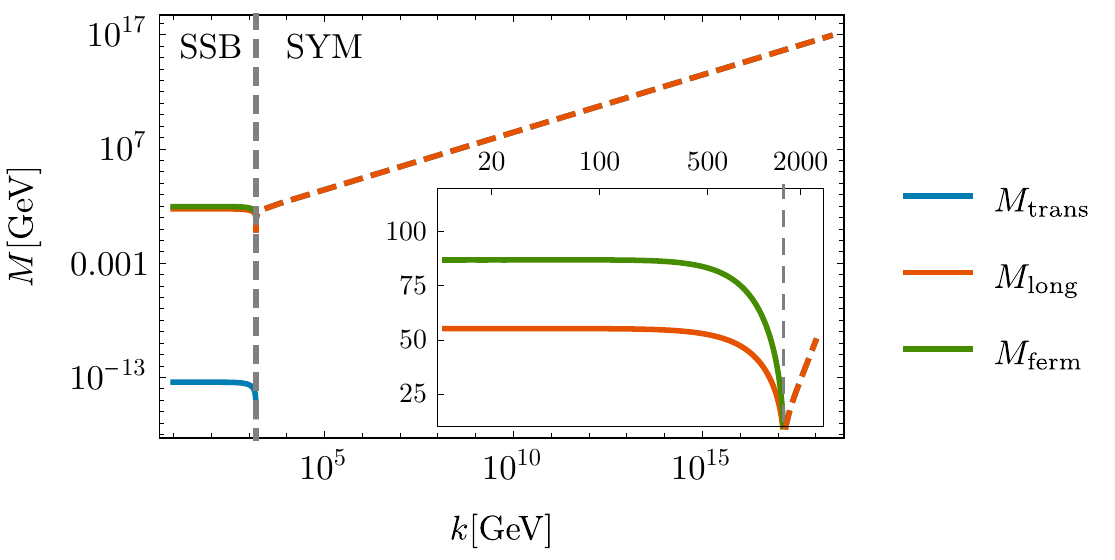}
    \caption{Masses of the transversal mode ($M_\text{trans}$), the longitudinal mode ($M_\text{long}$) and the fermion ($M_\text{ferm}$) as a function of the RG scale $k$. In the IR a huge mass ratio is generated naturally. 
    The UV initial conditions are $y(M_{\rm Planck})=0.4, \lambda_4(M_{\rm Planck})=0.001, z_n(M_{\rm Planck})=0.1$ and $\kappa(M_{\rm Planck})$ tuned such that $\bar{\kappa} = \left(246 \text{GeV}\right)^2$ in the IR.}
    \label{fig:masses_no_grav}
\end{figure}

\begin{figure}
    \centering
    \includegraphics{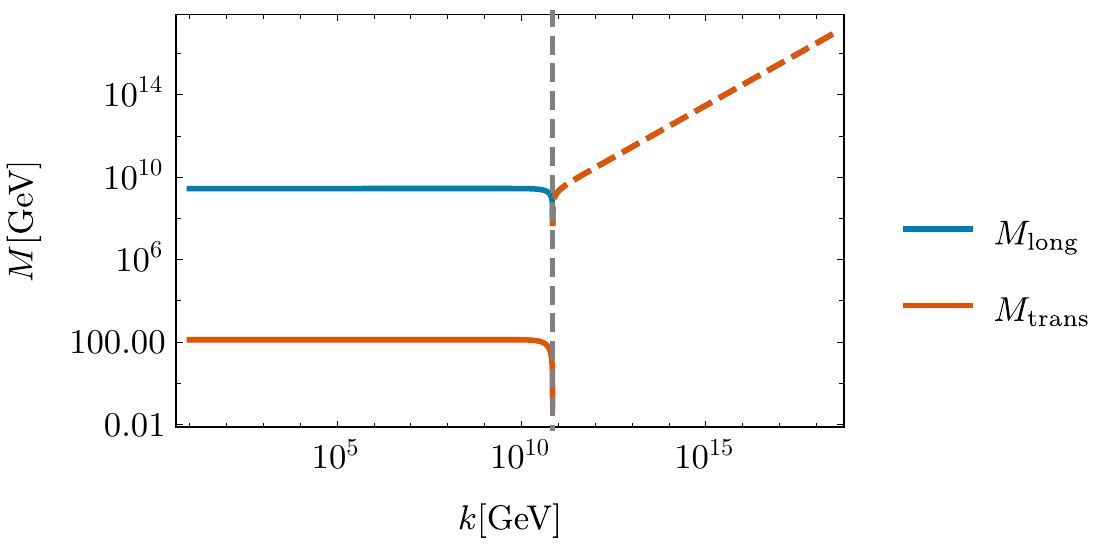}
    \caption{Masses of the transversal and the longitudinal mode as a function of the RG scale $k$. Here the UV initial condition for $\kappa$ is adjusted such that $M_\text{trans}$ is of the order of the Higgs mass. The other initial conditions are the same ones as in  Fig.~\ref{fig:masses_no_grav}.
    }
    \label{fig:transversal_higgs}
\end{figure}

For a concrete example of this mechanism, we set $n=6$, and assume that $k_\text{UV}$ is the Planck scale. We choose initial conditions for the couplings at $k_{\rm UV}$ to obtain the observed Higgs vacuum expectation value(vev) and  a fermion mass of the order of the top mass. The corresponding FRG flow equations for the symmetric and the symmetry-broken regime can be found in App.~\ref{app:floweqZn}.
The resulting flows for the couplings are displayed in Figs.~\ref{fig:deviation_from_crit}, \ref{fig:gamma_running} and \ref{fig:masses_no_grav}. 
As is apparent from these figures, $z_n$ flows towards tiny values in the IR, while $\kappa$ freezes out to the tuned vev, $\bar{\kappa}\ll M_{\rm Planck}$. 
Fig.~\ref{fig:masses_no_grav} illustrates that this generates large mass ratios as a result of the RG flow. In contrast, if $\kappa$ is not fine-tuned at $k_{\rm UV}= M_{\rm Planck}$, a vev of order  of the Planck scale is generated. In this case, one can still obtain a mass that is significantly smaller than the cutoff-scale, by tuning the $\z{6}$-symmetric coupling $z_n$, such that the transverse mass satisfies $M_{\rm trans}/M_{\rm Planck} \sim 10^{-17}$. We highlight that the tuning of $z_n$ required to obtain $M_{\rm trans}/M_{\rm Planck} \sim 10^{-17}$ is significantly less than the tuning of $\kappa$ to obtain $M_{\rm long}/M_{\rm Planck} \sim 10^{-17}$, cf.~Fig.~\ref{fig:transversal_higgs}. 

\section{Conclusions}

Symmetries determine our world, both at micro- and macroscales, yet, their fundamental origin remains a mystery.
In this paper, we have focused on two aspects of symmetries in fundamental physics.

First, we have explored the compatibility of global discrete $\z{n}$ symmetries with quantum gravity. 
We have found that quantum-gravity effects drive the corresponding couplings towards zero for $n \geq 4$. 
The effect is most pronounced in asymptotically safe gravity, where in a near-perturbative regime finite values of $\z{n}$-symmetric interactions cannot be achieved. 
The effect is less pronounced in an effective-field-theory setting for quantum gravity, where gravitational effects decrease the $\z{n}$ symmetric interactions, but nonzero values of the corresponding couplings can be achieved at and below the Planck scale, if the fundamental gravity theory features a breaking of continuous global symmetries. 

Second, we have explored the phenomenological implications of such a setting. 
As proposed in \cite{Leonard:2018sbi}, the presence of a $\z{n}$, $n>4$ symmetric term in an otherwise U(1) symmetric theory can naturally generate a mass hierarchy in the spontaneously symmetry-broken regime.
Here, we explore this idea for the first time in a four-dimensional Yukawa model and under the impact of quantum-gravity fluctuations. 
We discover that a transplanckian quantum-gravity regime with a finite UV cutoff can give rise to an RG flow that enters the symmetry-broken regime in the IR. 
There, the pseudo-Goldstone mode acquires a mass determined by the $\z{n}$ symmetric interaction. 
Since the latter is canonically irrelevant, that mass term is tiny compared to the mass of the standard massive mode. 
This mechanism is clearly highly interesting for phenomenology and model building beyond the Standard Model. 
It cannot be applied to scalars that interact via a gauge symmetry, as the latter would be violated explicitly by the $\z{n}$-symmetric interaction. 
Thus, it is not applicable to the Standard Model. 
On the other hand, beyond-Standard-Model sectors, e.g., dark sectors as well as extensions of the electroweak sector could feature mass hierarchies that find a natural explanation in the presence of a $\z{n}$ symmetry. For instance, a recent example in which a large mass hierarchy is required for a proposed solution to the Hubble tension, is given by \cite{Niedermann:2019olb}.

Our work paves the ground for future investigations targeted at the following questions:
\begin{enumerate}
\item[1)] Does the gravity-less, $\z{3}$-symmetric setting feature an interacting fixed point (cf. App. \ref{app:z3}), thereby circumventing the triviality problem of a $\phi^4$ scalar field theory in $d=4$?
 \item[2)] Does an asymptotically safe gravity-matter system give rise to an ``inverted" hierarchy, with the mass of the pseudo-Goldstone boson exceeding that of the standard massive mode within a $\z{3}$ symmetric setting?
 \item[3)] Does the interplay of two distinct scalar fields, charged differently under a global $\z{n}$ symmetry, provide a potential loophole to our no-go-hypothesis that excludes asymptotically safe models with $\z{n}$, $n\geq 4$ in the near-perturbative regime?
 \item[4)] Can the realization of discrete symmetries with $n\geq 4$ in asymptotically safe gravity also be excluded in the non-perturbative regime, or is the absence of discrete symmetries a consequence of the near-perturbative nature that asymptotically safe matter-gravity systems likely exhibit?
\end{enumerate}
Last but not least, an extension to a Standard-Model like setting, where the scalar field is charged under a gauge group, is of interest. 
In this case, the discrete symmetry cannot be imposed directly, as it is in conflict with gauge invariance.
However, it might be available as the remaining symmetry of a larger, spontaneously broken gauge group, see, e.g.,~\cite{Rachlin:2017rvm}.

In summary, in constructing QFTs, one a priori has a large amount of freedom in choosing the fundamental symmetries. 
In this paper we explore whether the interplay with quantum gravity can restrict  this freedom. 
Specifically, asymptotic safety could allow to severely constrain the space of possible matter theories. 
Its enhanced predictive power could fix coupling values within a single theory.  
Going beyond the values of interactions, asymptotic safety could even restrict the allowed symmetry structures by dividing them into a set of symmetries that feature non-trivial IR phenomenology and those that do not and can thus not be realized as fundamental symmetries.
In delineating the boundary between these two regions, our results indicate that some discrete symmetries remain trivial under the impact of quantum gravity.

\acknowledgments
A.~E.~is supported by a research grant (29405) from VILLUM FONDEN. M.~P.~is  supported by  a  scholarship  of  the  German  Academic  Scholarship Foundation and gratefully acknowledges the extended hospitality at CP3-Origins during various stages of this work.

\appendix

\section{Non-Perturbative Regime for \texorpdfstring{$n>3$}{n>3}}
\label{app:nonpert}

In Sec.~\ref{sec:single_field} we presented arguments why no fixed point for a $\z{n}$ symmetric regime exists in the near-perturbative regime. 
We focused on deformations of the Gau\ss{}ian fixed point.
Here, we explore if more general fixed points exist for $n>3$. 
We will first neglect gravitational effects, $f_s=0$, and then discuss the more general case $f_s \neq 0$.

The beta function for the coupling $z_n$ for $n>4$ schematically reads
\be
\label{eqn:beta_gn_app}
  \beta_{z_n} = \left(-4 + n - f_s + \# \lambda_4 \right)z_n + \# z_{n,2}
.\ee
Here $z_{n,2}$ is a coupling that belongs to a term with the structure $\phi \phi^\ast (\phi^n + (\phi^\ast)^n)$ and the various $\#$ represent various different numerical prefactors. 
A non-vanishing fixed point value for $z_n$ could arise from two contributions:
(i) The contribution proportional to $z_{n,2}$ can potentially shift the Gaussian fixed point. 
(ii) The interaction $z_n$ will induce a contribution to $\lambda_{2n-4}$, which (via various intermediate couplings in the U(1) sector) will feed back into $\lambda_4$, then inducing a higher order structure in $z_n$.

\begin{figure}
    \centering
    \includegraphics{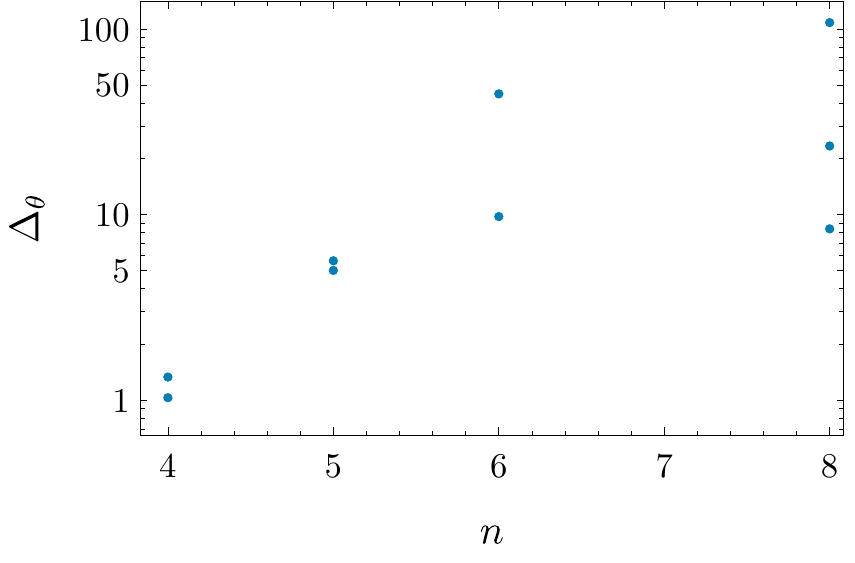}
    \caption{The measure $\Delta_\theta$ for all zeros of the beta functions with non-vanishing $z_n$ at varying $n$. For $n>4$ the resulting zeros of the beta-functions strongly deviate from canonical scaling. For $n=7$ we do not find any zero with $g_7\neq 0$.}
    \label{fig:non_perturbativity}
\end{figure}

We explore this possibility by computing the beta functions for a $\z{n}$ symmetric theory including all operators up to energy dimension $2n$. 
We then study the zeros of beta functions in these truncations. These zeros are only viable fixed-point candidates if the deviation from canonical scaling is not too large, otherwise our truncation scheme is invalidated.
For $n>4$ all zeros of the beta functions come with sizeable deviations from canonical scaling. 
We express this deviation in terms of the quantity
\be
    \Delta_\theta^2 = \frac{1}{N}\sum_i^N \left({\rm Re}[\theta_i] - d_i\right)^2,
\ee
where $\theta_i$ is the critical exponent, $d_i$ is the canonical dimension and the index $i$ runs over all couplings. 
Fig.~\ref{fig:non_perturbativity} illustrates that in a purely scalar system for $n>4$ any zero of the beta functions exhibits strong deviation from canonical scaling, indicating that it cannot reliably be interpreted as fixed points within our truncation. 

For $n=4$ a zero of the beta functions with $\order{1}$ deviations from canonical scaling appears. This is on the border to a non-perturbative regime and would require extended studies to confirm that indeed a scalar field theory could evade the triviality problem due to the existence of a $\z{4}$ symmetric interaction.

We conclude that without gravitational interactions within our truncations there is no indication for a stable fixed point for $n>3$.
Gravitational contributions will shift $z_n$ towards irrelevance and render all critical exponents smaller. 
This does not (significantly) reduce the departure from perturbativity.

In the non-perturbative regime our truncation (based on canonical power-counting) is expected to break down, our results are not self-consistent in that regime. 
Accordingly, our results cannot rule out a fixed point in the strongly interacting regime.

\section{Interacting Fixed Point for \texorpdfstring{$n=3$}{n=3}}\label{app:z3}

\begin{figure}
 \centering
 \includegraphics{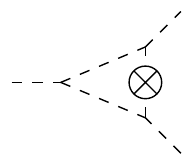}
 \caption{
 \label{fig:feynman_triangle}
 Triangle diagram that could induce a cubic contribution to the beta function. Diagrams with the regulator sitting on the other internal legs will also contribute.}
\end{figure}

To understand if an interacting fixed point can arise in the $\z{3}$ symmetric case, we focus on the beta function of $z_3$.
In $\beta_{z_3}$, a contribution $\order{z_3^3}$ could be expected to arise from the diagram depicted in Fig.~\ref{fig:feynman_triangle}. 
However, charge conservation at the vertex prevents this contribution from being nonzero.

A second potential source for a cubic term in $\beta_{z_3}$ is
the anomalous dimension. It reads
\be
  \eta_\phi = \frac{9 z_3^2}{8\pi^2 (1+\lambda_2)^4},
\ee
and enters the beta function for $z_3$ as
\be
  \beta_{z_3} = \frac{3}{2} z_3 \eta_\phi + \order{z_3^3},
\ee
hence inducing a cubic contribution in the beta function. 
This cubic contribution induces additional zeros of the beta function at non-vanishing values of $z_3$. 

For $f_s = 0$, these zeros come with large values for the quartic coupling $\lambda_4 \sim \order{20}$.
For non-vanishing gravitational contribution $f_s$, various of these additional zeros collide at $f_s \approx -0.4$. 
At $f_s = 0$ the critical exponents in a system with the three couplings $\lambda_2, z_3$ and $\lambda_4$ are $\theta_1 = 1.58, \theta_{2/3} = -0.54 \pm 1.58 i$.

\begin{figure}
    \centering
    \includegraphics[width=0.5\columnwidth]{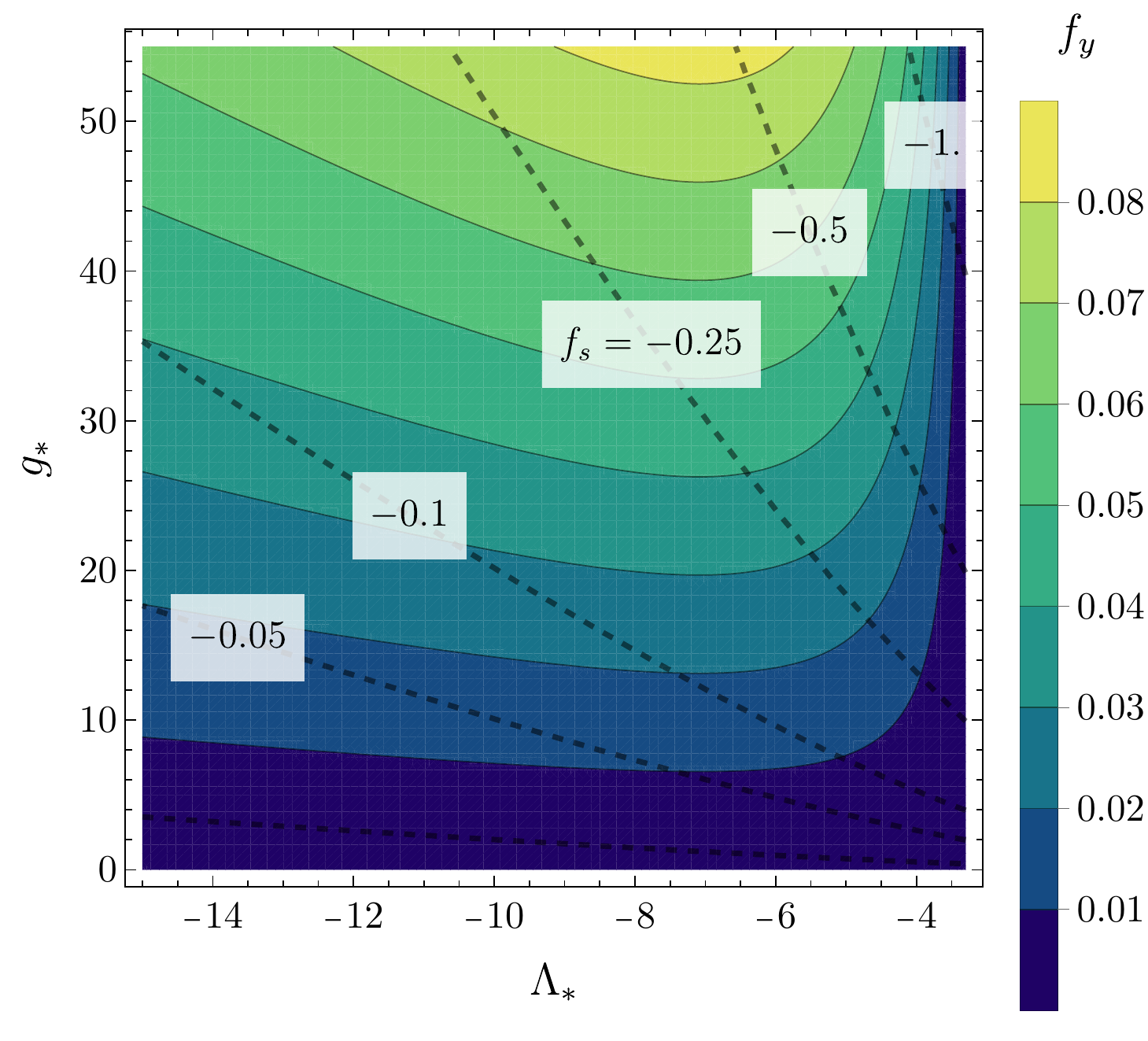}
    \caption{Gravitational contributions to the Yukawa coupling $f_y$ (color-coded) and scalar couplings $f_s$ (dashed lines) as a function of the (dimensionless) cosmological constant $\Lambda$ and Newton's constant $g$. Values of $\sim \order{1}$ for $f_s$ are only obtainable for large $g$.
    The expressions for $f_s$ and $f_y$ are taken from \cite{Eichhorn:2017ylw}.}
    \label{fig:fy_fs_lambda_g}
\end{figure}

In addition, for a small window in the vicinity of $f_s \approx -1$, a zero of the beta function exists. 
Indeed, in a truncation that considers all canonically relevant and marginal operators in the potential, for $f_y = 0.0025, f_s = -0.995$ we find the solution
\begin{align}
  \label{eq:fp_z3_value}
  y_{\ast} &= 0.23 & \lambda_{2\, \ast} &= 0.007 \\
  \lambda_{4\, \ast} &= -0.01 & z_{3\, \ast}&=0.16,
\end{align}
with a non-vanishing coupling $z_3$ and critical exponents
\begin{align}
  \theta_1 &= 1.02 & \theta_2 &= -0.0010 \\
  \theta_3 &= -0.0078 & \theta_4 &= -0.93
.\end{align}
Let us point out that the $\mathcal{O}(1)$ deviation of the three couplings in the scalar potential, $\lambda_2, \, \lambda_4$ and $z_3$ from their canonical scaling dimension is due to the gravity-induced anomalous dimension $f_s \approx -1$. In contrast, the small gravity-induced anomalous dimension for the Yukawa coupling, $f_y \approx 0$, results in the corresponding small shift of the corresponding scaling exponent from 0. These values for $f_s$ and $f_y$ are chosen freely to allow for an interacting fixed-point, irrespective of whether such values could be realized due to gravitational fluctuations. General studies of the gravitational parameter space indicate that these values are unlikely to be realized simultaneously \cite{Eichhorn:2017eht, deBrito:2019umw, Eichhorn:2020sbo}, see also Fig.~\ref{fig:fy_fs_lambda_g}.

To tentatively check if this zero is a genuine fixed point, we extend the truncation and include the order-6-couplings $\lambda_6$ and $z_6$. 
Both the fixed-point values and the critical exponents receive minor modifications under this extension of the truncation.

The relatively large shift in scaling dimensions due to $f_s$ causes $z_3$ to become irrelevant at this fixed point; the scalar mass remains as the only relevant parameter. Quite interestingly, one can therefore obtain a prediction for the IR value of this canonically relevant coupling.

\section{Two Fields}
\label{app:two_fields}
We consider the effective action 
\bea
  \label{eq:effective_action_two_field}
  \Gamma_k &=& \int \dd[4]{x} \sqrt{g} \Big( g^{\mu\nu}\partial_\mu \phi \partial_\nu \phi^\ast + g^{\mu\nu} \partial_\mu \chi \partial_\nu \chi^\ast + \sum_{i=0}^3 \sum_{j=0}^3 \frac{1}{i!\,j!}\lambda_{2i\,2j}(\phi \phi^\ast)^i (\chi \chi^\ast)^j k^{4-2i-2j}  \nonumber \\
  &&+ z_{03} k \left(\chi^3 + (\chi^\ast \right)^3) + z_{21} k \left((\phi^\ast)^2 \chi^\ast + \phi^2 \chi \right)   \nonumber \\
  &&+ z_{22} \left(\phi^2 (\chi^\ast)^2 + (\phi^\ast)^2 \chi^2\right)     \nonumber \\
  &&+ z_{41} k^{-1} \left( \phi^4 \chi^\ast + (\phi^\ast)^4 \chi \right) + z_{41a} k^{-1} \phi \phi^\ast \left( (\phi^\ast)^2 \chi^\ast + \phi^2 \chi \right) \nonumber \\
  &&+ z_{23} k^{-1} \chi \chi^\ast \left( \phi^2\chi+ (\phi^\ast)^2 \chi^\ast \right) + z_{23a} k^{-1} \phi \phi^\ast \left(\chi^3 + (\chi^\ast)^3 \right)     \nonumber \\
  &&+ z_{60}k^{-2} \left(\phi^6 + (\phi^\ast)^6\right) + z_{42} k^{-2} \left(\phi^2 \chi + (\phi^\ast)^2 \chi^\ast \right)^2   \nonumber \\
  &&+ z_{24} k^{-2} \left(\phi^2 \chi^4 + (\phi^\ast)^2 (\chi^\ast)^4 \right) + \frac{1}{2} z_{06} k^{-2} \left( \chi^3 + (\chi^\ast)^3\right)^2    \nonumber \\
  &&+ z_{42a} k^{-2} \phi \phi^\ast \left( \phi^2 (\chi^\ast)^2 + (\phi^\ast)^2 \chi^2 \right) + z_{24a} k^{-2} \chi \chi^\ast \left(\phi^2 (\chi^\ast)^2 + (\phi^\ast)^2 \chi^2 \right)
  \Big)
\eea
that contains all momentum-independent dimension six operators for two scalar fields $\phi$ and $\chi$ that are charged with charges $1$ and $-2$ under a common $\z{6}$ symmetry. By pairing up a field monomial and its complex conjugate, we ensure that the action remains real. \\
In addition, we include fermionic contributions of a Dirac fermion coupled to the $\phi$ field,
\be
  \Gamma_k^\text{ferm} =\!\int\!\dd[4]{x}\!\sqrt{g} \left( i \bar{\psi} \slashed{\nabla} \psi + y \left( \phi^\ast \bar{\psi}_R \psi_L\!-\! \phi \bar{\psi}_L \psi_R \right) \right)
,\ee
 following the conventions of \cite{Gies:2001nw,Gies:2009sv}.
Upon inclusion of gravitational contributions $f_s = -0.995$ and $f_y = 0.0025$ \footnote{Note that gravity distinguishes neither between the two scalars nor between the two fermions.}, we find a fixed point 
\begin{align}
\label{eq:fixed_point_two_field}
  z_{03} &= 0.15 & z_{21} &= 0.30 & z_{06}&=-5.0\cdot 10^{-7} \nonumber \\
  z_{22} &= -0.0026 & z_{23}&=0.00025 & z_{23a}&= 0.00017 \nonumber \\
  z_{24}&= -1.5 \cdot 10^{-6} & z_{24a} &= -0.0017 & z_{41} &= 2.1 \cdot 10^{-5}  \nonumber \\
  z_{41a} &= 9.4 \cdot 10^{-5} & z_{42} &= -7.7\cdot 10^{-7} & z_{42a} &= -0.00097  \nonumber \\
  z_{60} &= -2.6 \cdot 10^{-8} & \lambda_{02}&= 0.0077 & \lambda_{04} &= -0.012  \nonumber \\
  \lambda_{06} &= -0.0097 & \lambda_{20} &= 0.0052 & \lambda_{22} &= -0.011  \nonumber \\
  \lambda_{24} &= -0.012 & \lambda_{40}&=-0.0032 & \lambda_{42} &= -0.0064 \nonumber \\
  \lambda_{60} &= -0.0012 & y &= 0.24
\end{align}
with critical exponents
\begin{align}
  \label{eqn:app_two_fields_crit_exps}
  \theta_{1}&=1.02  & \theta_{2}&=1.00  & \theta_{3}&=-0.000611  \nonumber \\
  \theta_{4}&=-0.00466  & \theta_{5}&=-0.00786  & \theta_{6}&=-0.924  \nonumber \\ 
  \theta_{7}&=-0.946  & \theta_{8}&=-0.975  & \theta_{9}&=-1.00  \nonumber \\ 
  \theta_{10}&=-1.93  & \theta_{11}&=-1.96  & \theta_{12}&=-1.99  \nonumber \\ 
  \theta_{13}&=-2.00  & \theta_{14}&=-2.85  & \theta_{15}&=-2.87  \nonumber \\ 
  \theta_{16}&=-2.91  & \theta_{17}&=-2.95  & \theta_{18}&=-2.96  \nonumber \\ 
  \theta_{19}&=-3.00  & \theta_{20}&=-3.00  & \theta_{21}&=-3.00  \nonumber \\ 
  \theta_{22}&=-3.01  & \theta_{23}&=-3.01  
.\end{align}
In particular, that fixed point exhibits a finite fixed point value for both, the $\z{3}$ symmetric coupling $z_{03}$, as well as the portal coupling $z_{21}$, transferring the $U(1)$ violation from the $\chi$ to the $\phi$ sector. In such a setting one can induce a finite fixed point value for the coupling $z_{60}$ in the  $\z{6}$ symmetric $\phi$ sector via the effectively $\z{3}$ symmetric $\chi$ sector. Again the critical exponents are shifted with respect to the couplings' anomalous dimension by about $\order{1}$ due to the gravitational contribution $f_s$.

\section{FRG Flow Equations for \texorpdfstring{$\z{n}$}{Z\_n} Models}
\label{app:floweqZn}

In the following, we present the FRG flow equations for the $\z{n}$-symmetric model in the symmetric and in the symmetry-broken regime for $n \geq 6$, and $n$ even. 
Here, in both regimes we project by taking derivatives with respect to the parametrization in Sec.~\ref{sec:mass_hierarchy} $\rho := \phi \phi^\ast$ and $\tau := -\left(\phi^{n/2}-(\phi^\ast)^{n/2}\right)^2$. 
This is a slightly different projection than the one used in Sec.~\ref{sec:single_field}, where in the symmetric regime we project by simply taking derivatives with respect to $\phi$ and $\phi^\ast$. 
In the symmetric regime the beta functions for the couplings $\kappa, \lambda_4$ and $z_n$ read
\bea
    \beta_\kappa &=& - (2 + \eta_\phi) \kappa + \frac{(6-\eta_\phi)(1-6\kappa \lambda_4)}{48\pi^2 (1-\kappa \lambda_4)^3} +\frac{3 z_n (6-\eta_\phi) \kappa  \delta _{6,n}}{8 \pi ^2 \lambda_4 (1-\kappa  \lambda_4)^2}\\
    \beta_{\lambda_4} &=& 2 \eta_\phi \lambda_4 + \frac{5(6-\eta_\phi) \lambda_4^2}{48\pi^2 (1-\kappa \lambda_4)^3} -\frac{3 z_n (6-\eta_\phi)  \delta _{6,n}}{8 \pi ^2 (1-\kappa  \lambda_4)^2} \\ 
    \beta_{z_n} &=&(n-4+\frac{n}{2} \eta_\phi) z_n + \frac{z_nn (n-1) (6-\eta_\phi) \lambda_4}{96\pi^2 (1-\kappa \lambda_4)^3}
.\eea
Here $\delta_{6,n}$ is the Kronecker Delta. The corresponding term only contributes for $n=6$.
The direct quantum contributions to the beta function for the Yukawa coupling vanish
\be
    \beta_y = \left(\eta_\psi + \frac{1}{2} \eta_\phi\right) y 
.\ee
In the symmetry-broken regime the corresponding expressions are
\bea
    \beta_\kappa &=& - (2+\eta_\phi) \kappa + \frac{(6-\eta_\phi) z_n^2 n^4 \kappa^n}{64\pi^2 (1+2 \kappa \lambda_4)^2 (\kappa + z_n n^2 \kappa^{n/2})^2}  + \frac{(6-\eta_\phi) \kappa^2 (1+\kappa \lambda_4 + \kappa^2 \lambda_4^2)}{48\pi^2(1+2 \kappa \lambda_4)^2(\kappa + z_n n^2 \kappa^{n/2})^2} \nonumber \\
    & &- \frac{(6-\eta_\phi) z_n n^2 \kappa^{n/2} (2-4\kappa \lambda_4 + 8 \kappa^2 \lambda_4^2 - n (1+2\kappa \lambda_4)^2)}{384 \pi^2 \lambda_4 (1+2\kappa \lambda_4)^2 (\kappa + z_n n^2 \kappa^{n/2})^2} \\
    %%%%%%%%%%%%%%%%%%%%%%%%
    \beta_{\lambda_4} &=& 2 \eta_\phi \lambda_4 +
    \frac{\kappa ^6 \lambda _4^5 \left(6-\eta _{\phi }\right)}{12 \pi ^2 \left(1+2 \kappa  \lambda _4\right)^3 \left(\kappa+n^2 z_n \kappa ^{n/2} \right)^3}+\frac{\kappa ^5 \lambda _4^4 \left(6-\eta _{\phi }\right)}{8 \pi ^2 \left(1+2 \kappa  \lambda _4\right)^3 \left(\kappa+n^2 z_n \kappa ^{n/2} \right)^3}  \nonumber \\
    & &+\frac{\kappa ^4 \lambda _4^3 \left(6-\eta _{\phi }\right)}{16 \pi ^2 \left(1+2 \kappa  \lambda _4\right)^3 \left(\kappa+n^2 z_n \kappa ^{n/2} \right)^3}+\frac{5 \kappa ^3 \lambda _4^2 \left(6-\eta _{\phi }\right)}{48 \pi ^2 \left(1+2 \kappa  \lambda _4\right)^3 \left(\kappa+n^2 z_n \kappa ^{n/2} \right)^3}  \nonumber \\
    & &-\frac{n^2 z_n \left(6-\eta _{\phi }\right) \left(16 \kappa ^4 \lambda _4^4+32 \kappa ^3 \lambda _4^3-3 \kappa ^2 \lambda _4^2+8 \kappa  \lambda _4+1\right) \kappa ^{n/2}}{96 \pi ^2 \left(1+2 \kappa  \lambda _4\right)^3 \left(\kappa+n^2 z_n \kappa ^{n/2} \right)^3}+\frac{n^6 z_n^2 \left(6-\eta _{\phi }\right) \kappa ^{n-1}}{768 \pi ^2 \left(\kappa+n^2 z_n \kappa ^{n/2} \right)^3}  \nonumber \\
    & &+\frac{3 \lambda _4^2 n^6 z_n^3 \left(6-\eta _{\phi }\right) \kappa ^{\frac{3 n}{2}}}{32 \pi ^2 \left(1+2 \kappa  \lambda _4\right)^3 \left(\kappa+n^2 z_n \kappa ^{n/2} \right)^3}-\frac{n^5 z_n^2 \left(6-\eta _{\phi }\right) \kappa ^{n-1}}{384 \pi ^2 \left(\kappa+n^2 z_n \kappa ^{n/2} \right)^3} -\frac{n^4 z_n \left(6-\eta _{\phi }\right) \kappa ^{n/2}}{768 \pi ^2 \left(\kappa+n^2 z_n \kappa ^{n/2} \right)^3} \nonumber \\
    & &+\frac{9 \lambda _4^2 n^4 z_n^2 \left(6-\eta _{\phi }\right) \kappa ^{n+1}}{32 \pi ^2 \left(1+2 \kappa  \lambda _4\right)^3 \left(\kappa+n^2 z_n \kappa ^{n/2} \right)^3}+\frac{n^3 z_n \left(6-\eta _{\phi }\right) \left(4 \kappa  \lambda _4+3\right) \kappa ^{n/2}}{384 \pi ^2 \left(\kappa+n^2 z_n \kappa ^{n/2} \right)^3}  \\
    %%%%%%%%%%%%%%%%%%%%%%%%%%%%%
    \beta_{z_n} &=& (n-4 + \frac{n}{2}\eta_\phi) z_n + \frac{z_n^2 (6-\eta_\phi) n^2 (n-1) \kappa^{n/2}(n-2+2(n-1)\kappa \lambda_4)}{192 \pi^2 (1+2\kappa \lambda_4)^2 (\kappa + z_n n^2 \kappa^{n/2})^2}  \nonumber \\ 
    & &+ \frac{z_n (6-\eta_\phi)n(n-1) \kappa^2 \lambda_4 (1+\kappa \lambda_4)}{96\pi^2 (1+2\kappa \lambda_4)^2 (\kappa + z_n n^2 \kappa^{n/2})^2}   + \frac{z_n^3 (6-\eta_\phi) n^4 (n^2-3n+2) \kappa^{n-1} }{384\pi^2(1+2\kappa \lambda_4)^2 ( \kappa + z_n n^2 \kappa^{n/2})^2}
\eea
The Yukawa beta function reads
\bea
    \beta_y &=& \left(\eta_\psi + \frac{\eta_\phi}{2}\right) y -\frac{y^3}{32 \pi ^2 \left(1 + \kappa y^2\right)^2 \left(1 + 2 \kappa  \lambda _4\right)}+\frac{y^3}{16 \pi ^2 \left(1 + \kappa y^2\right)^3 \left(1 + 2 \kappa  \lambda _4\right)}  \nonumber \\
    & &+ \frac{3 n^2 y^3 z_n \lambda _4 \kappa ^{\frac{n}{2}+2}}{8 \pi ^2 \left(1 + \kappa y^2\right) \left(\kappa+n^2 z_n \kappa ^{n/2} \right)^2 \left(1 + 2 \kappa  \lambda _4\right)^2} -\frac{y^3 \kappa }{16 \pi ^2 \left(1 + \kappa y^2\right)^3 \left(\kappa+n^2 z_n \kappa ^{n/2} \right)} \nonumber \\
    & & +\frac{3 y^3 \lambda _4 \kappa ^3}{8 \pi ^2 \left(1 + \kappa y^2\right) \left(\kappa+n^2 z_n \kappa ^{n/2} \right)^2 \left(1 + 2 \kappa  \lambda _4\right)^2} +\frac{y^3 \left((n-2) n^2 z_n \kappa ^{n/2}+2 \lambda _4 \kappa ^2\right) \kappa ^2}{16 \pi ^2 \left(1 + \kappa y^2\right) \left(\kappa+n^2 z_n \kappa ^{n/2} \right)^3}  \nonumber \\
    & &+\frac{y^5 \left(n^2 z_n \kappa ^{n/2}+2 \left(1+\kappa  \lambda _4\right) \kappa \right) \kappa ^2}{32 \pi ^2 \left(1 + \kappa y^2\right)^2 \left(\kappa+n^2 z_n \kappa ^{n/2} \right)^2 \left(1 + 2 \kappa  \lambda _4\right)} -\frac{y^5 \kappa }{16 \pi ^2 \left(1 + \kappa y^2\right)^3 \left(1 + 2 \kappa  \lambda _4\right)} \nonumber \\
    & &+\frac{y^5 \kappa ^2}{16 \pi ^2 \left(1 + \kappa y^2\right)^3 \left(\kappa+n^2 z_n \kappa ^{n/2} \right)}-\frac{y^3 \lambda _4 \left((n-2) n^2 z_n \kappa ^{n/2}+2 \lambda _4 \kappa ^2\right) \kappa ^2}{8 \pi ^2 \left(1 + \kappa y^2\right) \left(\kappa+n^2 z_n \kappa ^{n/2} \right)^2 \left(1 + 2 \kappa  \lambda _4\right)^2}  \nonumber \\
    & &-\frac{3 y^3 \lambda _4 \kappa ^2}{8 \pi^2 \left(1 + \kappa y^2\right) \left(\kappa+n^2 z_n \kappa ^{n/2} \right) \left(1 + 2 \kappa  \lambda_4 \right)^2}+\frac{y^3 \left((n-2) n^2 z_n \kappa ^{n/2}+2 \lambda _4 \kappa ^2\right) \kappa }{32 \pi ^2 \left(1 + \kappa y^2\right)^2 \left(\kappa+n^2 z_n \kappa ^{n/2} \right)^2}  \nonumber \\
    & &+\frac{y^3 \left((n-2) n^2 z_n \kappa ^{n/2}+2 \lambda _4 \kappa ^2\right) \kappa }{16 \pi ^2 \left(1 + \kappa y^2\right) \left(\kappa+n^2 z_n \kappa ^{n/2} \right)^2 \left(1 + 2 \kappa  \lambda _4\right)}+\frac{y^3 \kappa }{32 \pi ^2 \left(1 + \kappa y^2\right)^2 \left(\kappa+n^2 z_n \kappa ^{n/2} \right)}  \nonumber \\
    & &-\frac{y^3 \left(n^2 z_n \kappa ^{n/2}+2 \left(\kappa  \lambda _4+1\right) \kappa \right) \kappa }{32 \pi ^2 \left(1 + \kappa y^2\right)^2 \left(\kappa+n^2 z_n \kappa ^{n/2} \right)^2 \left(1 + 2 \kappa  \lambda _4\right)}-\frac{3 y^3 \lambda _4 \kappa }{16 \pi ^2 \left(1 + \kappa y^2\right)^2 \left(1 + 2 \kappa  \lambda _4\right)^2} \nonumber \\
    & &-\frac{y^5 \left(n^2 z_n \kappa ^{n/2}+2 \left(\kappa  \lambda _4+1\right) \kappa \right) \kappa }{32 \pi ^2 \left(1 + \kappa y^2\right)^2 \left(\kappa+n^2 z_n \kappa ^{n/2} \right) \left(1 + 2 \kappa  \lambda _4\right)^2} -\frac{3 y^3 \lambda _4 \kappa }{8 \pi ^2 \left(1 + \kappa y^2\right) \left(1 + 2 \kappa  \lambda _4\right)^3} \nonumber \\
    & &-\frac{y^3 \left((n-2) n^2 z_n \kappa ^{n/2}+2 \lambda _4 \kappa ^2\right) \kappa }{16 \pi ^2 \left(1 + \kappa y^2\right) \left(\kappa+n^2 z_n \kappa ^{n/2} \right)^2 \left(1 + 2 \kappa  \lambda _4\right)^2}  \nonumber \\
    & &+\frac{y^3 \left(n^2 z_n \kappa ^{n/2}+2 \left(\kappa  \lambda _4+1\right) \kappa \right)}{32 \pi ^2 \left(1 + \kappa y^2\right)^2 \left(\kappa+n^2 z_n \kappa ^{n/2} \right) \left(1 + 2 \kappa  \lambda _4\right)^2} 
,\eea
where in the last expression we already dropped the anomalous dimensions in the numerators.
\bibliography{references}
\end{document}